\documentclass[11pt,sort&compress]{elsarticle}

\input{preamble.tex}
\usetikzlibrary{shapes.arrows, patterns, calc}
\usepackage{tikz-3dplot}
\usepackage{bbm}
\ifpdf
  \DeclareGraphicsExtensions{.eps,.pdf,.png,.jpg}
\else
  \DeclareGraphicsExtensions{.eps}
\fi

\usepgfplotslibrary{groupplots}
\usetikzlibrary{arrows}
\usetikzlibrary{shapes}
\usetikzlibrary{decorations.text}
\usetikzlibrary{arrows.meta}
\usetikzlibrary{calc}
\usetikzlibrary{fit}

\definecolor{steelblue}{HTML}{A1BDC7}
\definecolor{orange}{HTML}{D98C21}
\definecolor{silver}{HTML}{B0ABA8}
\definecolor{rust}{HTML}{B8420F}
\definecolor{seagreen}{HTML}{2E6B69}
\definecolor{joshua}{HTML}{FBDC7F}
\definecolor{darksky}{HTML}{154c79}

\colorlet{lightsilver}{silver!30!white}
\colorlet{darkorange}{orange!85!black}
\colorlet{darksilver}{silver!85!black}
\colorlet{darksteelblue}{steelblue!85!black}
\colorlet{darkrust}{rust!85!black}
\colorlet{darkseagreen}{seagreen!85!black}

\colorlet{RYB1}{rust}
\colorlet{RYB2}{seagreen}
\colorlet{RYB3}{orange}
\colorlet{RYB4}{purp}
\colorlet{RYB5}{steelblue}
\colorlet{RYB6}{joshua}
\colorlet{RYB7}{darksky}

\pgfplotscreateplotcyclelist{newcolors}{
{RYB1,every mark/.append style={fill=RYB1,mark size={2.5}},mark=*},
{RYB2,every mark/.append style={fill=RYB2},mark=square*},
{RYB3,every mark/.append style={fill=RYB3,mark size={3}},mark=triangle*},
{RYB4,every mark/.append style={fill=RYB4,mark size={3}},mark=diamond*},
{RYB5,every mark/.append style={fill=RYB5,mark size={3}},mark=pentagon*},
{RYB6,every mark/.append style={fill=RYB6,mark size={4}},mark=10-pointed star},
{RYB7,every mark/.append style={fill=RYB7},mark=*},
}

\pgfplotsset{
    compat=1.18,
    standard/.style={
        scale only axis,
        width=0.5\textwidth,
        enlarge x limits=0.05,
        enlarge y limits=0.05,
        max space between ticks=40,
    },
    every tick label/.append style={font=\footnotesize},
    every legend/.append style={font=\small},
    every node/.append style={font=\small},
    every axis label/.append style={font=\small},
    cycle list name=newcolors,
}

\tikzset{
    every node/.append style={font=\small},	
}

\usepackage{mathtools}
\usepackage{stackengine}
\usepackage{fixmath}
\mathtoolsset{centercolon}  

\renewcommand{\phi}{\varphi}
\renewcommand{\epsilon}{\varepsilon}

\newcommand{\cnst}[1]{\mathrm{#1}}

\newcommand{\zerovct}{\bm{0}} 
\newcommand{\Id}{\mathbf{I}} 

\newcommand{\zeromtx}{\bm{0}}

\newcommand{\vct}[1]{\mathbold{#1}}

\newcommand{\tp}{{\mathsf{T}}}

\newcommand{\trace}{\operatorname{tr}}

\newcommand{\diff}{\Phi}

\DeclarePairedDelimiterX{\infdivx}[2]{(}{)}{%
  #1\;\delimsize\|\;#2%
}

\newcommand{\divergence}{\operatorname{div}}

\begin{document}

\hypersetup{
  linkcolor=darkrust,
  citecolor=seagreen,
  urlcolor=darkrust,
  pdfauthor=author,
}

\begin{frontmatter}

\title{
Shocks without shock capturing: Information geometric regularization of \\ finite volume methods for Navier--Stokes-like problems
}

\author[1]{Anand Radhakrishnan}
\author[1]{Benjamin Wilfong}
\author[1,2,3]{Spencer H.\ Bryngelson\corref{cor1}}
\ead{shb@gatech.edu}
\author[4]{Florian Sch\"afer\corref{cor1}}
\ead{florian.schaefer@nyu.edu}

\address[1]{School of Computational Science \& Engineering, Georgia Institute of Technology, Atlanta, GA 30332, USA \vspace{-0.15cm}}
\address[2]{Daniel Guggenheim School of Aerospace Engineering, Georgia Institute of Technology, Atlanta, GA 30332, USA \vspace{-0.15cm}}
\address[3]{George W.\ Woodruff School of Mechanical Engineering, Georgia Institute of Technology, Atlanta, GA 30332, USA \vspace{-0.15cm}}
\address[4]{Courant Institute of Mathematical Sciences, New York University, New York, NY 10012, USA}

\cortext[cor1]{Equal contribution}

\date{}

\end{frontmatter}

\begin{abstract}
    Shock waves in high-speed fluid dynamics produce near-discontinuities in the fluid momentum, density, and energy.
    Most contemporary works use artificial viscosity or limiters as numerical mitigation of the Gibbs--Runge oscillations that result from traditional numerics.
    These approaches face a delicate balance in achieving sufficiently regular solutions without dissipating fine-scale features, such as turbulence or acoustics.
    Recent work by Cao and Sch\"afer 
    introduces information geometric regularization (IGR), the first inviscid regularization method for fluid dynamics. 
    IGR replaces shock singularities with smooth profiles of adjustable width, without dissipating fine-scale features. 
    This work provides a strategy for the practical use of IGR in finite-volume-based numerical methods.
    We illustrate its performance on canonical test problems and compare it against established approaches based on limiters and Riemann solvers.
    Results show that the finite volume IGR approach recovers the expected solutions in all cases.
    Across canonical benchmarks, IGR achieves accuracy competitive with WENO and LAD shock-capturing schemes in both smooth and discontinuous flow regimes.
    The IGR approach is computationally light, with meaningfully fewer memory accesses and arithmetic operations per time step.
\end{abstract}

\blfootnote{All code available at \url{https://github.com/sbryngelson/figr} (multidimensional results using MFC) and \url{https://github.com/f-t-s/IGR_Manual_1dFVM} (unidimensional results in Julia).}

\section{Introduction}

\subsection*{The problem: Shock waves in compressible CFD}

Shock waves pose a central challenge in the numerical simulation of compressible fluids, particularly in high-speed fluid dynamics.
Such waves introduce near-discontinuities in the momentum, density, and energy fields.
These discontinuities cause Gibbs--Runge oscillations for numerical methods based on higher-order polynomial reconstruction~\citep{shu1998advanced,gottlieb1997gibbs,canuto2006spectral}. 
Tailoring the mesh to the shock can overcome this problem, but it is prohibitively complicated for geometrically complex shock-interaction problems.
Alternatively, the field of \emph{shock capturing} aims to coarse-grain the effects of the microscopically thin shock without resolving it or interfering with the solution in other regions of the flow.

\subsection*{Shock capturing: Localized viscosity and limiters}

Fluid dynamics studies relying on computation of compressible flow often add some form of localized artificial viscosity near the shock using so-called shock sensors~\cite{vonneumann1950method,puppo2004numerical,cook2005hyperviscosity,fiorina2007artificial,mani2009suitability,barter2010shock,guermond2011entropy,bruno2022fc,dolejvsi2003some}. 
This expands the shock to the grid scale, mitigating Gibbs--Runge oscillations.
However, these viscous approaches must balance regularity against the preservation of fine-scale features, which often have physical meaning.
Further,  approaches based on local viscosity can break down at high Mach numbers~\cite{glaubitz2019smooth}.

Instead of regularizing the Navier--Stokes equations with viscosity, limiters lower the approximation order near shocks~\cite{van1979towards,ray2018artificial,liu1994weighted,harten1997uniformly,shu1998advanced,guermond2018second,kuzmin2020monolithic}. 
Limiters are often more robust than artificial diffusivity, but likewise need to balance the taming of shocks with the preservation of fine-scale features. 
Furthermore, they are intrusive changes to the numerical methods that can prohibit certain downstream applications.
In particular, the nonlinear switching logic inherent to limiters limiters render adjoint-based sensitivity computation undefined or adds systematic biases~\cite{lozano2019watch,bodony2022adjoint}, often requiring practitioners to accept the errors due to the ``freezing'' of the limiters \cite{nemec2006aerodynamic}.
IGR's entropic pressure, being the solution of a linear self-adjoint elliptic problem, does not share this drawback.

\subsection*{Information geometric regularization (IGR): The first inviscid shock regularization}

Recent work by \citet{cao2023information} introduces the first regularization of compressible CFD that does not rely on viscosity.
They first derive this \emph{information geometric regularization} (IGR) in the pressureless inviscid (Euler) case, and observe that shock formation amounts to the collision and merging of particle trajectories.
They then change the geometry of particle motion such that trajectories cannot cross, but instead asymptotically approach each other (see \cref{fig:regularization}). 
This approach avoids the formation of grid-scale singularities while preserving post-shock behavior.

When forces due to pressure and viscosity are included, IGR yields a regularized conservation law that replaces singular shocks with grid-scale smooth profiles at a user-defined scale.
The higher-order grid-scale smoothness of these profiles yields robust solutions even at high Mach numbers (see \citet{wilfong2025simulating} for an application to a two-species Mach 14 flow).
Because of its inviscid nature, this regularization has little impact on fine-scale features, such as turbulence or acoustic waves.
In the unidimensional pressureless case, the entropic pressure equation reduces to $\Sigma / \rho - \alpha\,\partial_{x}\left(\Sigma / \rho\right) = 2\alpha\,(\partial_x u)^2$ and the regularized momentum equation reads $\partial_t (\rho u) + \partial_x(\rho u^2 + \Sigma) = 0$. 
This setting forms the starting point for the derivation of IGR by \citet{cao2023information}.
In the unidimensional pressureless case, \citet{cao2024information} rigorously establish the existence of global smooth IGR solutions and their convergence to entropy solutions in the limit of vanishing regularization.
Recent work of \citet{barham2025hamiltonian} investigates the connection between IGR and Hamilton regularizations as proposed by \citet{guelmame2022hamiltonian}.

\subsection*{Outline}
This work presents a finite-volume IGR implementation and illustrates its performance against canonical test problems.
It is organized as follows.
We begin in \cref{s:shock_formation} by reviewing shock formation in the compressible Navier--Stokes equations and placing IGR in the context of viscous and limiter-based shock capturing.
\Cref{s:igr} introduces the IGR target equations and briefly sketches the geometric motivation.
We highlight the practical bottlenecks addressed by IGR in \cref{s:fvm}, reviewing finite-volume methods and the roles of nonlinear reconstructions and Riemann solvers.
Our finite-volume discretization of the IGR entropic pressure and the resulting algorithm are developed in \cref{s:igr_fvm}.
Numerical results are presented in \cref{section:results}.
\Cref{s:discussion} closes with comparisons to related work, limitations, and conclusions.

\section{Shock formation in the compressible Navier--Stokes equations}\label{s:shock_formation}

\subsection{Compressible Navier--Stokes equations}\label{s:NS}

The compressible Navier--Stokes equations are
\begin{align}
    \frac{\partial \rho}{\partial t} + \nabla \cdot (\rho \bu) &= 0, \label{e:mass}\\
    \frac{\partial (\rho \bu)}{\partial t} + \nabla \cdot (\rho \bu \bu^{\tp} + p \bI - \boldsymbol\tau)  &= 0, \label{e:momentum}\\
    \frac{\partial (E)}{\partial t} + \nabla \cdot \left[ (E + p) \bu - \boldsymbol{\tau} \cdot \bu \right] &= 0,\label{e:energy}
\end{align}
which represent the conservation of mass, momentum, and energy.
Their variables include pressure $p$, density $\rho$, total energy $E$, time $t$, velocity $\bu = \{u_1,u_2,u_3\}$ (in the 3D case); $\bI$ is the identity tensor.
To simplify the presentation, we use the ideal gas law equation of state
\begin{equation}
    \label{eqn:state_eqn}
    p = (\gamma - 1)\rho e, 
    \quad \text{for} \quad 
    e = E/\rho - \|\bu\|^2 / 2,
\end{equation}
where $\gamma$ is the ratio of specific heats (adiabatic exponent).
However, the method is not restricted to this choice.
The coefficients $\mu$ and $\zeta$ are the shear and bulk viscosity, and the viscous stress tensor is
\begin{equation}
    \tau_{ij} = \mu\left(\frac{\partial u_i}{\partial x_j} + \frac{\partial u_j}{\partial x_i}\right) + \left(\zeta - {2}\mu/{3}\right) \delta_{ij} \frac{\partial u_k}{\partial x_k},
    \label{eqn:constitutive_law}
\end{equation}
where $\delta_{ij}$ is the Kronecker delta, $x_i$ are the coordinate directions and $i,j \in \{1,2,3\}$ for the 3D case.

\subsection{Shock formation}

The nonlinear terms of the compressible Navier--Stokes equations cause wavefronts to steepen; faster-moving regions overtake slower ones, producing shocks (near-discontinuities) in pressure, density, and velocity.
Viscosity smooths these features, but physical shocks remain thin (typically of the order of the mean free path~\cite{koreeda1995front}), far below typical Navier--Stokes grid resolution scales (and, indeed, the range of validity of the equations themselves).
Thus, in the case of the Navier--Stokes equations, some degree of additional treatment via modeling or numerics is required to represent shock behavior faithfully.
These additions modify the governing equations or the numerical methods to coarse-grain the shock to practical grid scales, a procedure known as shock capturing.

\subsection{Viscous PDE regularizations}

The most basic approach to regularizing the Navier--Stokes equations for shock handling is to add small viscous terms~\cite{vonneumann1950method}, modifying \cref{e:mass,e:momentum,e:energy} as
\begin{align}
    \frac{\partial \rho}{\partial t} + \nabla \cdot (\rho \bu) &= \epsilon_{\rho} \nabla^2 \rho \\
    \frac{\partial (\rho \bu)}{\partial t} + \nabla \cdot (\rho \bu \bu + p \bI - \boldsymbol\tau) &= \epsilon_{\rho \bu} \nabla^2 (\rho \bu) \\
    \frac{\partial (E)}{\partial t} + \nabla \cdot \left[ (E + p) \bu - \boldsymbol{\tau} \cdot \bu \right] &= \epsilon_{E} \nabla^2 (E),
\end{align}
where $\epsilon$ is the artificial viscosity coefficient. 
The value of $\epsilon$ determines the trade-off between shock capturing and the preservation of other flow features, like acoustics.
More sophisticated approaches make this trade-off more palatable by letting $\epsilon$ depend on the current-time flow solution.
The resulting nonlinear viscosity acts as a \emph{shock sensor}, localizing it to the vicinity of the shock.

A notable viscous approach is localized artificial diffusivity (LAD)~\cite{cook2005hyperviscosity,mani2009suitability}, which regularizes the Navier--Stokes equations as
\begin{align}
    \frac{\partial \rho}{\partial t} + \nabla \cdot (\rho \bu) &= 0 \\
    \frac{\partial (\rho \bu)}{\partial t} + \nabla \cdot (\rho \bu \bu^{\tp} + p \bI - \boldsymbol\tau) &= \nabla (\zeta^* \nabla \cdot \bu) \\
    \frac{\partial (E)}{\partial t} + \nabla \cdot \left[ (E + p) \bu - \boldsymbol{\tau} \cdot \bu \right] &= \nabla (\zeta^* \nabla \cdot \bu) \cdot \bu.
\end{align}
Rather than adding a viscous regularization to each component of the conservation law, it modifies the equation by locally increasing its bulk viscosity. 
It is modulated by a function of the local flow state that detects under-resolved features (such as shocks), a so-called shock sensor.
Using artificial bulk (as opposed to shear) viscosity $\zeta^*$ helps restrict additional dissipation to shocks rather than to the vortical fine-scale structures typical of turbulence.
Many variants exist, but a representative example proposed by \citet{mani2009suitability} is to choose 
$\zeta^* \propto \overline{\rho \max\left(0, -\nabla \cdot \bu\right)}$, where the overline indicates the application of Gaussian smoothing with length scale commensurate to the grid spacing.  
The constant of proportionality is taken to be of the order of the squared grid spacing.

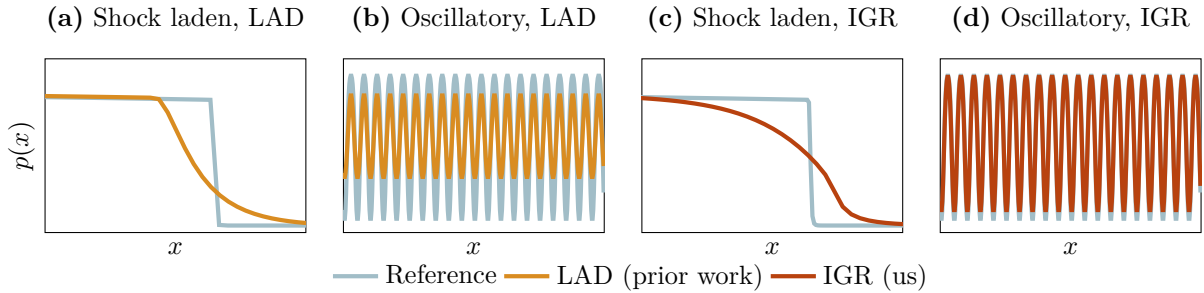
\begin{figure}[tbph]
    \centering
    \tikzsetnextfilename{igr_lad}
    \begin{tikzpicture}\end{tikzpicture}
    \caption{Comparison of localized artificial diffusivity (LAD) and information geometric regularization (IGR) approaches for shock treatment (modified from \cite{wilfong2025simulating}, with author permission.)}
    \label{fig:lad_igr}
\end{figure}

\subsection{A discrete alternative: Limiters}\label{s:limiters}

Rather than regularizing the PDE, an alternative approach changes the numerical discretization to handle discontinuities.
ENO schemes~\cite{harten1987uniform} adaptively select the smoothest stencil at each cell interface.
WENO~\cite{liu1994weighted,shu1998advanced} generalizes this by forming a convex combination of candidate stencils, weighted by smoothness indicators.
MUSCL~\cite{vanleer1979towards,vanleer1974towards} reconstructs piecewise-linear profiles within each cell.
All three families achieve high-order accuracy in smooth regions while suppressing spurious oscillations near shocks.

These and related Total Variation Diminishing (TVD) schemes employ limiters, that is, nonlinear functions (for example, minmod, superbee, van Leer, MC, or the stencil selection in (W)ENO) that reduce the reconstruction order near discontinuities.
The choice of limiter governs the trade-off between numerical dissipation and shock sharpness: more diffusive limiters, like minmod, are robust but smear shocks, and less diffusive ones, like superbee, preserve sharper shocks but are less stable.
A promising alternative is convex limiting~\cite{guermond2018second,kuzmin2020monolithic}, which applies the minimum correction necessary to preserve invariant domain properties of the PDE.

\section{Information geometric regularization}\label{s:igr}

\subsection{The IGR target equations}

The information geometric regularization of \cref{e:mass,e:momentum,e:energy} complements the physical pressure $p$ with the \emph{entropic pressure} $\Sigma$~\cite{cao2023information}, defined by the auxiliary elliptic problem
\begin{equation}
    \rho^{-1} \Sigma - \alpha \divergence(\rho^{-1} \nabla \Sigma) = \alpha  \left(\trace^2\left([\jac]\right) + \trace\left([\jac]^2\right) \right),
    \label{e:entropic_pressure}
\end{equation}
where $\jac$ is the flow Jacobian matrix with components $(\jacel)_{ij} \coloneqq \partial_{j} u_i$ and $\alpha > 0$ is the regularization parameter controlling the width of the smoothed shock profile.
This yields the information geometric regularization 
\begin{align}
    \frac{\partial \rho}{\partial t} + \nabla \cdot (\rho \bu) &= 0, \label{e:igr_mass}\\
    \frac{\partial (\rho \bu)}{\partial t} + \nabla \cdot (\rho \bu \bu^{\tp} + (p + \Sigma) \bI - \boldsymbol\tau) &= \zerovct, \label{e:igr_momentum} \\
    \frac{\partial (E)}{\partial t} + \nabla \cdot \left[ (E + (p + \Sigma)) \bu - \boldsymbol{\tau} \cdot \bu \right] &= 0.
    \label{e:igr_energy}
\end{align}

\subsection{Inviscid regularization with smooth solutions}

\Cref{e:igr_mass,e:igr_momentum,e:igr_energy} have solutions that replace shocks with profiles that are smooth at the scale of $\sqrt{\alpha}$.
At the same time, these profiles propagate and interact with other flow structures and with each other as if they were nominal shock regularized by vanishing viscosity.
The absence of viscous regularization means that fine-scale flow features, such as acoustic or entropy waves and turbulence features, are preserved.

IGR does not act as a diffusive mechanism in the conventional sense.
Rather than damping the solution over time, it adds a pressure-like term that prevents the trajectories of gas volumes from colliding.
The elliptic source $\trace^2(\jac) + \trace(\jac^2)$ is largest wherever velocity gradients are sharpest, so $\Sigma$ grows in those regions and adds to the pressure that opposes the trajectory crossing, as illustrated in Figure 2.
As described in \citet[Section 7.1]{cao2023information}, IGR can be interpreted as a ``nonlocal and sign-indefinite viscosity''.
The nonlocality of $\Sigma$ enables IGR to slow down gas particles approaching the shock \emph{before they reach it}, thus enabling IGR to produce the higher-order smooth profiles shown in \cref{fig:lad_igr}.
The sign-indefinite nature of IGR ensures that it only activates in systematically compressive flows (shocks), but not in oscillatory features due to sound waves or turbulence, as illustrated in \cref{fig:lad_igr}. 
This is the distinction from viscous regularizations, which damp all flow gradients regardless of their character.

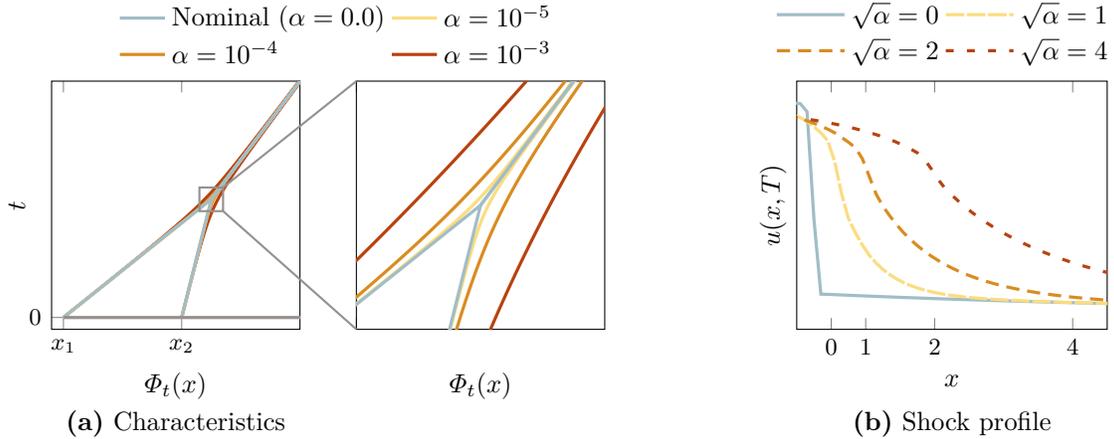
\begin{figure}[tbph]
    \centering
    \tikzsetnextfilename{regularization}
    \begin{tikzpicture}\end{tikzpicture} 
    \caption{
    \textbf{Information geometric regularization (IGR).}
    In the pressureless case, IGR regularizes particle trajectories $t \mapsto \diff_t(x)$ such that instead of colliding, particles asymptotically approach each other ((a), modified from \cite{cao2023information}, with author permission).
    With pressure, this replaces the singular shocks with smooth profiles on a scale $\sqrt{\alpha}$~(b).
    In the $\alpha \to 0^+$ limit, one recovers the Euler equations, and the auxiliary IGR equation \cref{e:entropic_pressure} has the trivial solution $\Sigma \equiv 0$.
    }
    \label{fig:regularization}
\end{figure}

\subsection{A primer on the geometric derivation of IGR}

We denote as $\diff_t$ the \emph{flow map}, which tracks where each fluid parcel, initially at position $\vct{x}$, moves to at time $t$.
The flow map $\diff_t$ satisfies $\vct{u}(\diff_t(\vct{x})) = \dot{\diff}_t(x)$ and $\rho(\diff_t(x)) = \det [\cnst{D} \diff_t(x)]$, where $\cnst{D} \diff_t$ denotes the deformation gradient of the flow map, distinct from the velocity gradient $\jac$ in \cref{e:entropic_pressure}.
Solutions to \cref{e:mass,e:momentum} describe a time evolution on the manifold of possible flow maps~\cite{arnold1966geometrie,khesin2021geometric}.

In the pressureless and inviscid case ($p = 0, \vct{\tau} = \zeromtx$), $t \mapsto \diff_t$ describes a straight line on the manifold of diffeomorphisms (differentiably invertible maps).
Shocks form when $\diff_t$ reaches the boundary of this manifold, meaning that two points $\vct{x}_1 \neq \vct{x}_2$ are mapped to the same $\diff_t(\vct{x}_1) = \diff_t(\vct{x}_2)$.
After shock formation, the trajectories $t \mapsto \diff_t(\vct{x}_1)$ and $t \mapsto \diff_t(\vct{x}_2)$ merge; $\diff_t$ evolves on the boundary of the diffeomorphism manifold.
\citet{cao2023information} regularize this equation by changing the notion of ``straight lines'' on the diffeomorphism manifold such that they instead asymptotically approach the boundary, but do not reach it. 

With IGR, trajectories $t \mapsto \diff_t(\vct{x})$ merge asymptotically, without ever colliding, as shown in \cref{fig:regularization}.
This avoids the formation of singularities while preserving the long-time behavior.
When adding back the pressure to the resulting equations, their solutions replace singular shocks with smooth profiles on a scale determined by the parameter $\alpha$ that measures the strength of the regularization.

\subsection{IGR in practice}

As outlined by \citet{cao2023information}, a computational promise of IGR is obviating the need for nonlinear numerical methods, like limiters or Riemann solvers, for representing shock behavior.
IGR produces no regularization in pure shear: the flow Jacobian there is traceless and nilpotent, so both $\trace^2(\jac)$ and $\trace(\jac^2)$ on the right-hand side of \cref{e:entropic_pressure} vanish and $\Sigma = 0$.
IGR thus targets compressive features without affecting shear-dominated structures.
IGR is formulated to reduce the numerical artifacts encountered by these approaches, such as artificial dissipation or carbuncles. 
At the same time, replacing nonlinear limiters and stencil-selection logic with standard linear reconstruction and flux evaluation reduces the computational cost of simulating flows with shocks, as demonstrated by \citet{wilfong2025simulating}.
Note that the full IGR scheme is nonlinear: the nonlinearity enters through the entropic pressure $\Sigma$, whose elliptic right-hand side depends quadratically on the velocity gradient.
This work presents IGR in the same finite-volume framework used by traditional approaches to compressible flows.
We next review this framework and identify the bottlenecks that IGR resolves.

\section{Finite Volume Methods}\label{s:fvm}

Finite volume methods (FVMs) discretize the computational domain into control volumes and store cell-averaged solution values.
Conservation is enforced by balancing fluxes across cell faces via the divergence theorem, making FVMs well-suited for problems involving shocks and other discontinuous fields.

\subsection{Fundamental concept}

In a finite volume formulation, the conservation law $\partial_t \phi + \nabla \cdot \mathbf{F}=S$ is integrated over each control volume $\Omega$.
Applying the divergence theorem converts the volume integral of the divergence into a surface integral of outward-pointing face fluxes, yielding the semi-discrete balance
\begin{gather}
    \frac{\dd}{\dd t}\int_{\Omega} \phi \, \dd\Omega = S_{\Omega} + \sum_{f\in \{\text{faces of $\Omega$}\}} F_f\,A_f,
\end{gather}
where $F_f$ is the average flux through the face $f$, and $A_f$ is that face's area.
High-order reconstruction reduces the mismatch between left- and right-hand interface states, decreasing the dissipation introduced by the Riemann solver.
Some Riemann solvers (for example, Lax--Friedrichs) entail more dissipation, encouraging numerical stability, while others (for example, HLLC, discussed next) are more aggressive.

\subsection{Discrete finite volume formulation}\label{s:lax}

We describe a typical finite-volume solver of the one-dimensional compressible Euler equations. 
The same strategy extends directly to the 3D compressible Navier--Stokes equations on a Cartesian grid by successively computing fluxes in each coordinate direction.
One stores each conserved variable as a cell average, evaluates numerical fluxes at every cell face, and applies the 1D update formula along each coordinate direction.

The 1D compressible Euler equations can be expressed as a conservation law,
\begin{gather}
    \frac{\partial \bU(x,t)}{\partial t} + 
    \frac{\partial}{\partial x}\bF \left(\bU(x,t)\right) = \bzero,
    \label{eq:euler-1d}
\end{gather}
with conservative state variables $\bU$ and fluxes $\bF(\bU)$ as
\begin{gather}
    \bU =
    \begin{pmatrix}
        \rho \\ \rho u \\ E
    \end{pmatrix}
    \quad
    \text{and}
    \quad
    \bF(\bU) =
    \begin{pmatrix}
        \rho u \\ \rho u^{2}+p \\ (E+p)u
    \end{pmatrix},
    \label{eq:U_F}
\end{gather}
$I_i=[x_{i-1/2},x_{i+1/2}]$ is a grid cell of width $\Delta x$ and the cell average at time $t^n$ is
\begin{gather}
    \bU_i^{n}= \frac{1}{\Delta x}\int_{I_i} \bU(x,t^n) \dd x.
    \label{e:integral}
\end{gather}
Evaluating \cref{e:integral} and time integrating \cref{eq:euler-1d} requires the fluxes at the finite volume faces $\bF(\bU_\mathrm{L, R})$.
The pressure depends on the state variables through the equation of state. 
Here, we consider the case of polytropic gas laws for simplicity, with an equation of state of the form $p = (\gamma - 1) \left(E - \rho u^2 / 2 \right)$ with adiabatic exponent $\gamma$.

The standard first-order scheme for computing the fluxes at the half-cell distances uses piecewise constant reconstruction, with resulting left and right interface states $\bU_{i+1/2}^{\mathrm{L}}=\bU_i$ and $\bU_{i+1/2}^{\mathrm{R}}=\bU_{i+1}$.
To achieve higher-order accuracy, we replace these constants by values obtained from a polynomial reconstruction of degree $5$ constructed from neighboring cell averages and then evaluate the polynomial at the cell faces.
For example, a fifth-order accurate polynomial reconstruction uses the reconstruction weights
\begin{gather}
    \label{e:reconstruction}
    w^\mathrm{L} = \{ 2/60, -13/60, 47/60, 27/60, -3/60 \}, \\
    w^\mathrm{R} = \{ -3/60, 27/60, 47/60, -13/60, 2/60 \}, 
\end{gather}
and defines the reconstructions as
\begin{gather}
    \bU_{i+1/2}^{\mathrm{L}} = \sum \limits_{j = - 2}^{+ 2} w_j^\mathrm{L} \bU_{i+j}, \qquad 
    \bU_{i+1/2}^{\mathrm{R}} = \sum \limits_{j = - 1}^{+ 3} w_j^\mathrm{R} \bU_{i+j}.
    \label{e:reconstruction_formula}
\end{gather}
The numerical method's spatial order of accuracy matches the accuracy of this polynomial reconstruction.

The numerical fluxes are denoted $\widehat \bF_{i\pm 1/2}$, and their evaluation depends on the numerical method used, including the Riemann solver.
For this work, we only fully express the local Lax--Friedrichs (LF) Riemann solver, which is also known as the Rusanov numerical flux.
With $\bU_{\mathrm{L}}=\bU_{i+1/2}^{\mathrm{L}}$ and $\bU_{\mathrm{R}}=\bU_{i+1/2}^{\mathrm{R}}$, the LF numerical flux is
\begin{gather}
    \widehat{\bF}_{i+1/2} = 
        \frac{1}{2}
        \left[ \bF \left(\bU_{\mathrm{L}}\right) + 
        \bF \left(\bU_{\mathrm{R}}\right) \right] - \frac{\lambda_{i+1/2}}{2}\left( \bU_{\mathrm{R}} - \bU_{\mathrm{L}} \right),
    \label{eq:flux-lf}
\end{gather}
where
\begin{gather}
    \lambda_{i+1/2}= \max
    \{|u_{\mathrm{L}}| + c_{\mathrm{L}},
      |u_{\mathrm{R}}| + c_{\mathrm{R}}\}
    \label{eq:alpha-lf}
\end{gather}
is the largest local wave speed, $u$ is the velocity, and $c$ is the sound speed.
Integrating \cref{eq:euler-1d} over $I_i$ and the time interval $t^n$ to $t^{n+1}$ gives the update equation for the state variables:
\begin{gather}
    \bU_i^{n+1}= \bU_i^{n} - \frac{\Delta t}{\Delta x}\left(\widehat \bF_{i+1/2}^{n} - \widehat \bF_{i-1/2}^{n}\right).
    \label{eq:fv-update}
\end{gather}

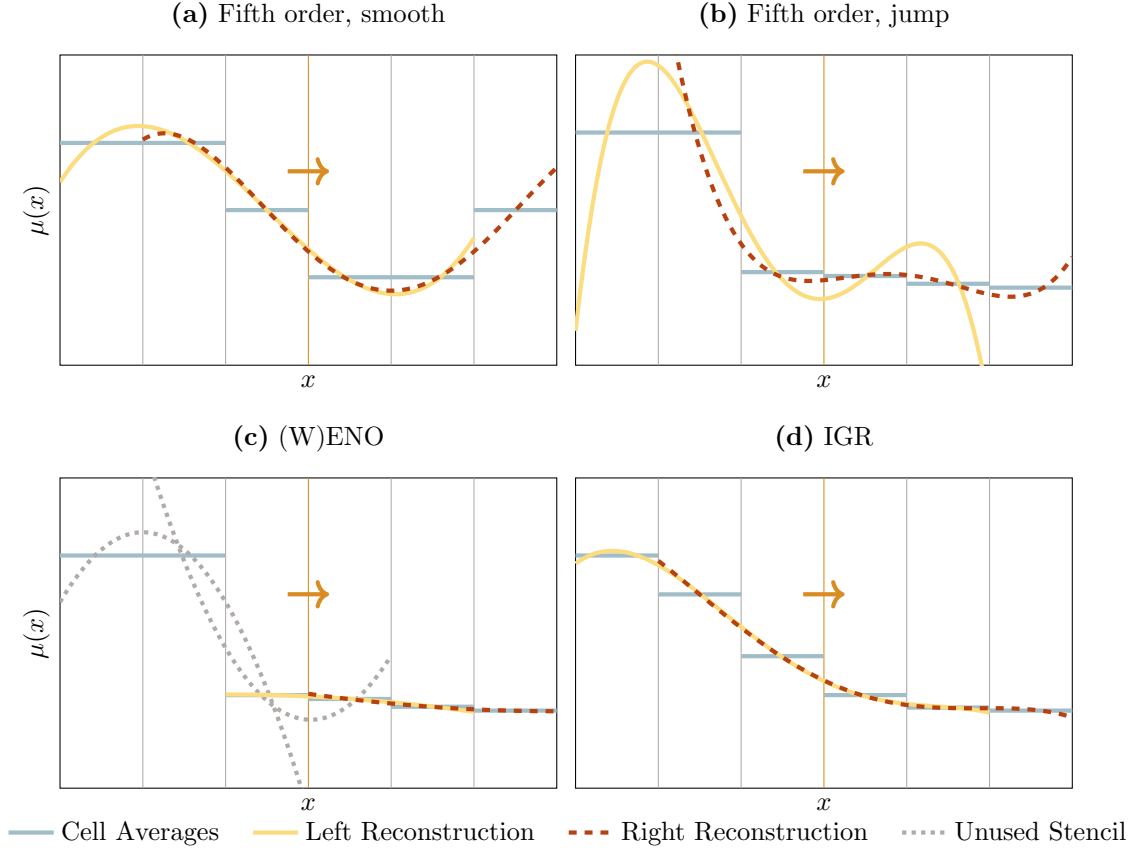
\begin{figure}
    \centering
    \tikzsetnextfilename{fvm}
    \begin{tikzpicture}\end{tikzpicture}
    \caption{Comparison of high-order polynomial reconstruction strategies in FVM near a jump discontinuity (orange).
    Blue horizontal segments are cell averages; the orange arrow marks the target face.
    (a)~Fifth-order reconstruction from smooth data yields a single consistent polynomial.
    (b)~The same stencil near a jump produces oscillatory left/right reconstructions.
    (c)~(W)ENO selects among candidate sub-stencils (dotted), preferring the smoothest one.
    (d)~IGR smooths the cell averages themselves, enabling a single high-order reconstruction without nonlinear stencil selection.}
    \label{fig:FVWENOIGR}
\end{figure}

This numerical scheme is stable so long as the Courant--Friedrichs--Lewy (CFL) condition is satisfied:
\begin{gather}
    0 < \frac{\Delta t}{\Delta x}\max_i \lambda_{i+1/2} \le 1.
    \label{eq:cfl-lf}
\end{gather}
The diffusive term in \cref{eq:flux-lf} is designed to recover a method sufficiently stable for transport-dominated problems. 
In smooth flow fields, the difference $\left( \bU_{\mathrm{R}} - \bU_{\mathrm{L}} \right)$ vanishes as the fifth power of the mesh width, limiting the spurious dissipation in this setting.

In low-regularity situations, such as shock formation, high-order reconstructions cause oscillations due to Gibbs--Runge phenomena. 
Treating discontinuous behavior has heretofore required limiters such as ENO~\cite{harten1987uniform} and MUSCL~\cite{vanleer1974towards,vanleer1979towards} schemes that locally reduce the approximation order, or other viscous mechanisms.
The resulting lower order reconstructions increase the difference $\left( \bU_{\mathrm{R}} - \bU_{\mathrm{L}} \right)$ and thus the diffusive effects of Lax-Friedrichs Riemann solvers, as illustrated in \cref{fig:FVWENOIGR}.

Mitigating these effects requires the use of sophisticated approximate Riemann solvers to compute fluxes at cell interfaces, such as 
Harten--Lax--van Leer (HLL) and HLLC~\citep{harten1983upstream,toro1994restoration,toro2009riemann}.
Those also involve nonlinear computations and intermediate variables, such as wave speeds.
Together, the nonlinear computation of interface reconstructions and Riemann solvers requires that large temporary arrays be stored in device memory.
Furthermore, they prevent the fusion of compute kernels that could otherwise compute these quantities in a single kernel pass.
The confluence of these features prevents the aggregation of compute kernels that update grid-level quantities (such as fluxes) in place, resulting in computation and memory overhead.
In contrast, the discrete flow Jacobian required by IGR uses a compact 3-point stencil and adds no temporary storage; the dominant savings come from eliminating the nonlinear stencil selection, smoothness indicators, and characteristic decompositions of WENO, not from avoiding gradient computation per se.
Quantitative performance comparisons are provided in \citet{wilfong2025simulating}.
Beyond these computational issues, the reliance on grid-aligned treatments used by these Riemann solver-driven approaches is suspected to give rise to the carbuncle phenomenon \cite{chauvat2005shock}.
Linear schemes, such as the Lax--Friedrichs method discussed in \cref{s:lax}, avoid this shortcoming.

IGR replaces shocks with smooth solutions \cite{cao2024information}. 
As illustrated in \cref{fig:FVWENOIGR}, this enables the use of higher-order reconstructions throughout, and thus foregoes nonlinear numerics in the form of limiters or Riemann solvers. 

\section{Solving the IGR equations with finite volume methods}\label{s:igr_fvm}

Empirical and theoretical evidence suggest that the information geometrically regularized \cref{e:igr_mass,e:igr_momentum,e:igr_energy} produce solutions that are smooth on the scale $\sqrt{\alpha}$ (\cref{fig:regularization}, \cite{cao2023information,cao2024information}).
Thus, higher-order polynomial reconstructions avoid the formation of Gibbs--Runge oscillations for a mesh width $\lesssim \sqrt{\alpha}$. 
As a result, no limiters are needed; the difference between left and right reconstructions vanishes to high order in the mesh width, and the Lax--Friedrichs scheme is only weakly diffusive.

\subsection{Discretization of the entropic pressure}

The only change needed for these benefits is the addition of the entropic pressure $\Sigma$ to the physical pressure $p$.
Since the regularization term is only of order $\alpha \approx \Delta x^2$, we find that a piecewise constant approximation of $\Sigma$ yields satisfactory results in practice, which are shown later in \cref{section:results}. 
Denoting $\rho_{ijk}$ as the average density and $\rho\vct{u}_{ijk}$ as the average momentum over the $ijk$-th cell, and letting $\Delta_x$, $\Delta_y$, and $\Delta_z$ denote the grid spacings in the $x$, $y$, and $z$ coordinate directions, we define the discrete flow Jacobian $\jac_{ijk}$ at the $ijk$-th cell using the central stencil
\begin{equation}
    \label{e:discrete_jacobian}
    \jac_{ijk} 
    = 
    \begin{pmatrix}
        \frac{1}{2 \Delta_x} \left(\frac{\rho\vct{u}_{(i+1)jk}}{\rho_{(i+1)jk}} - \frac{\rho \vct{u}_{(i-1)jk}}{\rho_{(i-1)jk}}\right)
        & \frac{1}{2 \Delta_y} \left(\frac{\rho\vct{u}_{i(j+1)k}}{\rho_{i(j+1)k}} - \frac{\rho\vct{u}_{i(j-1)k}}{\rho_{i(j-1)k}}\right)
        & \frac{1}{2 \Delta_z} \left(\frac{\rho\vct{u}_{ij(k+1)}}{\rho_{ij(k+1)}} - \frac{\rho\vct{u}_{ij(k-1)}}{\rho_{ij(k-1)}}\right)
    \end{pmatrix},
\end{equation}
which is, again, a linear and grid-local operation.

Solving the elliptic equation \cref{e:entropic_pressure} for $\Sigma$ using the finite difference discretization of the Laplacian yields
\begin{equation}
    \label{e:discrete_elliptic}
    \begin{split}
        \frac{\Sigma_{ijk}}{\rho_{ijk}} 
        &+ \alpha \Bigg(
        \sum \limits_{|i - \tilde{i}| = 1} \frac{2\left(\Sigma_{ijk} - \Sigma_{\tilde{i}jk}  \right)}{\Delta_x^2\left( \rho_{\tilde{i}jk} + \rho_{ijk}  \right)} 
        + 
        \sum \limits_{|j - \tilde{j}| = 1} \frac{2\left(\Sigma_{ijk} - \Sigma_{i\tilde{j}k}  \right)}{\Delta_y^2\left( \rho_{i\tilde{j}k} + \rho_{ijk}  \right)} 
        +
        \sum \limits_{|k - \tilde{k}| = 1} \frac{2\left(\Sigma_{ijk} - \Sigma_{ij\tilde{k}}  \right)}{\Delta_z^2\left( \rho_{ij\tilde{k}} + \rho_{ijk}  \right)}
        \Bigg)\\
        &= \alpha \left( \trace^2\left(\jac_{ijk}\right) + \trace\left(\jac_{ijk}^2\right) \right)
    \end{split}
\end{equation}
Numerous methods exist for solving this system of equations. 
One choice is Jacobi iteration, which amounts to computing the next iterate $\Sigma_{ijk}^{n + 1}$ from the previous one $\Sigma_{ijk}^{n}$ using the formula
\begin{equation}
    \label{e:jacobi}
    \begin{split}
        \Sigma_{ijk}^{(n + 1)} =
        \frac{
        \alpha \left( \trace^2\left(\jac_{ijk}\right) + \trace\left(\jac_{ijk}^2\right) \right)
        + \alpha \left(
        \sum \limits_{|i - \tilde{i}| = 1}   \frac{2\Sigma_{\tilde{i}jk}^{(n)}/\Delta_x^2}{ \rho_{\tilde{i}jk} + \rho_{ijk} }
        + \sum \limits_{|j - \tilde{j}| = 1}   \frac{2\Sigma_{i\tilde{j}k}^{(n)}/\Delta_y^2}{ \rho_{i\tilde{j}k} + \rho_{ijk} }
        + \sum \limits_{|k - \tilde{k}| = 1}   \frac{2\Sigma_{ij\tilde{k}}^{(n)}/\Delta_z^2}{ \rho_{ij\tilde{k}} + \rho_{ijk}}
        \right)
        }{
        \frac{1}{\rho_{ijk}}
        + \alpha \left(
        \sum \limits_{|i - \tilde{i}| = 1}   \frac{2/\Delta_x^2}{ \rho_{\tilde{i}jk} + \rho_{ijk} }
        + \sum \limits_{|j - \tilde{j}| = 1}   \frac{2/\Delta_y^2}{ \rho_{i\tilde{j}k} + \rho_{ijk} }
        + \sum \limits_{|k - \tilde{k}| = 1}   \frac{2/\Delta_z^2}{ \rho_{ij\tilde{k}} + \rho_{ijk}}
        \right)
        }
    \end{split}.
\end{equation}

Jacobi iteration is an effective iterative method for solving the elliptic equation \cref{e:discrete_elliptic}, and is shown to be sufficient to compete with established numerical methods in \cref{section:results}.
Other iterative methods, such as Gauss--Seidel or successive over-relaxation (SOR), can further accelerate this step.
In our experience, the elliptic solve is not the computational bottleneck.
After the first time step, the iteration is warm-started from the previous time step's $\Sigma$, so very few iterations are required to converge.
Combined with the small value of $\alpha \propto \Delta x^2$ and the absence of global communication, the solve is well-suited to GPU execution.
Quantitative cost comparisons with WENO are provided in \citet{wilfong2025simulating}.

\subsection{Boundary conditions for $\Sigma$}

The elliptic problem that defines $\Sigma$ requires boundary conditions. 
We use Neumann boundary conditions on $\Sigma$, enforcing $\partial_n \Sigma = 0$ at the wall.
This is the boundary-normal component of $\divergence\left(\Sigma \Id\right)$, so $\Sigma$ contributes no spurious normal force at the wall.
Equivalently, the wall acts as a mirror plane for $\Sigma$, which is the correct condition for a shock propagating parallel to a flat wall.
Algorithmically, this restricts the sums of \cref{e:discrete_elliptic,e:jacobi} to $1 \leq \tilde{i} \leq m_x, 1 \leq \tilde{j} \leq m_y, 1 \leq \tilde{k} \leq m_z$, where $m_x$, $m_y$, and $m_z$ are the number of grid cells in each coordinate direction.
The exception to this choice is rotation-symmetric or periodic boundary conditions.
There, the elliptic problem uses the same wrap-around boundary conditions as the overall problem. 

\subsection{Smoothing boundary and initial conditions}
\label{sec:smoothing}

Shock treatment via IGR is based on the principle that solutions of the regularized equations are as grid-scale smooth as their initial conditions. 
For such grid-scale smooth initial conditions, one recovers grid-scale smooth solutions always, mitigating Gibbs--Runge oscillations in higher-order reconstructions. 
However, many canonical problems in fluid dynamics, such as shock tubes, have discontinuous initial or boundary conditions. 
Without limiters or viscous regularizations, discontinuities of initial or boundary conditions will trigger Gibbs--Runge oscillations that pollute the simulation. 

This problem can be addressed by applying a small amount of smoothing to ensure smoothness \emph{on the grid scale} of the initial and boundary conditions. 
In practice, one can apply a small, constant number of diffusive pseudo-time steps to the state variables $\bU$ in the initial and boundary conditions to achieve this. 
IGR preserves the smoothness of the initial and boundary conditions at the grid scale.
Thus, different from the filtering steps in spectral methods, \emph{no further smoothing needs to be applied during the simulation.}

\subsection{Discretization of the resulting equation}

The above-described steps ensure that the PDE solutions remain smooth, enabling the use of higher-order accurate reconstructions, such as the fifth-order accurate ones in \cref{e:reconstruction}. 
The high-order accuracy of the reconstructions allows using the straightforward Lax--Friedrichs Riemann solver while incurring little numerical diffusion.
Because IGR solutions are smooth at the scale $\sqrt{\alpha}$, the left/right interface mismatch satisfies $\|\bU_R - \bU_L\| = O((\Delta x  / \sqrt{\alpha})^5)$. Thus, the LF dissipation term is also $O((\Delta x  / \sqrt{\alpha})^5)$ and can be controlled by an appropriate choice of $\sqrt{\alpha}$.
Numerous time integrators are available for the resulting semi-discretization. 
Strong-stability-preserving Runge--Kutta schemes perform well in practice~\cite{gottlieb2001strong}; we use them for the numerical experiments in \cref{section:results}.

\section{Numerical results}\label{section:results}

\subsection{Convergence}

Using PDE-based regularizations, such as IGR, raises three distinct questions about the convergence of the resulting numerical method. 
The first question is whether the numerical solution converges as $\Delta x \rightarrow 0$ for a fixed $\alpha$ (and thus, a fixed PDE) to the continuous solution of the regularized PDE.
The second is whether the exact solution of the regularized PDE converges as the regularization parameter tends to zero ($\alpha \rightarrow 0$).
Third, and most practically relevant, is the convergence of the numerical approximation of IGR as both the regularization parameter and the grid scale tend to zero, following their natural scaling relationship ($\sqrt{\alpha} \propto \Delta x \rightarrow 0$).

\begin{figure}[tbph]
    \centering
    \tikzsetnextfilename{igr_convergence}
    \begin{tikzpicture}\end{tikzpicture}
    \caption{We investigate the convergence of IGR and its numerical solution in three different regimes. 
    In (a), we fix $\alpha$ and observe the convergence with decreasing grid size. 
    We see (near) second-order convergence both before and after the shock forms.
    In (b), we observe the convergence of IGR solutions with decreasing $\alpha$, computed on a fine grid. 
    We observe second-order (in $\sqrt{\alpha}$) convergence before shock formation, and first-order convergence thereafter.
    In (c), we investigate the behavior as both $\alpha$ and $\Delta x$ tend to zero. 
    We choose the scaling relationship $\sqrt{\alpha} \propto \Delta x$, which results in shock widths proportional to the grid size, as would be expected in a practical application of our approach. We again observe second-order convergence before and first-order convergence after shock formation.}
    \label{fig:igr_convergence}
\end{figure}
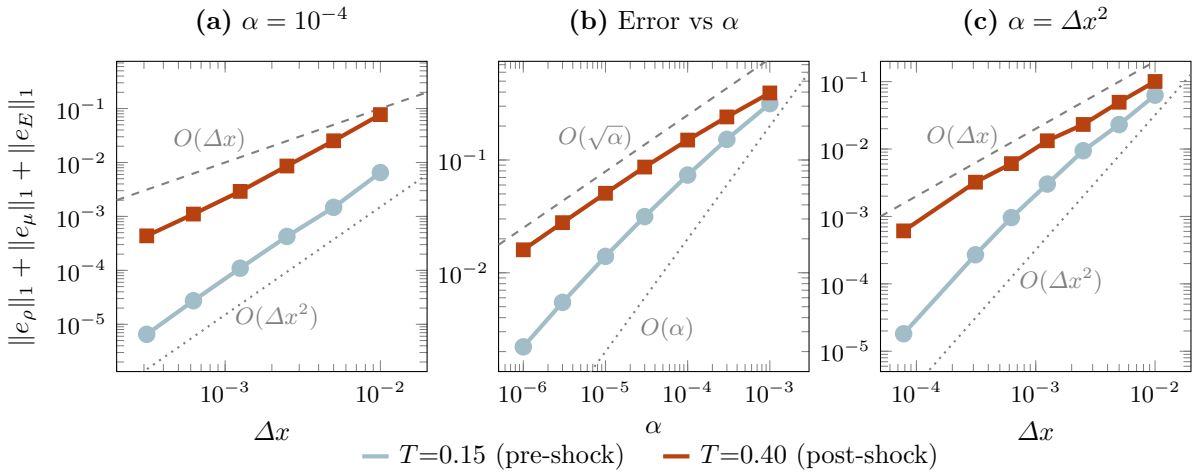

\Cref{fig:igr_convergence} demonstrates convergence in these three regimes on a one-dimensional test case with initial condition $u(x) = 1.5 \sin(2 \pi x), \rho(x) \equiv 1, e(x) = 4$, distinguishing behavior before and after shock formation.
For a fixed $\alpha$, we observe second-order convergence before shock formation and almost second-order convergence after shock formation. 
Since we are using a second-order accurate discretization of the entropic pressure $\Sigma$, this matches the expected order of accuracy. 
When decreasing $\alpha$ for a fixed and sufficiently small $\Delta x$, we observe second-order (in $\sqrt{\alpha}$) convergence before shock formation and first-order convergence after a shock has formed. 
The former is to be expected because the entropic pressure is of the order $\alpha (\partial_x u)^2$. 
The first-order convergence in the shock arises because in a shock of width $\sqrt{\alpha}$, the flow derivative $\partial_x u$ is of the order $\alpha^{-1/2}$, and thus the entropic pressure is constant to leading order. 
Thus, the overall decrease of the $L^1$ norm of the entropic pressure is only due to the decreasing width of the shock, resulting in first-order convergence. 
Similar behavior is observed in virtually all approaches to shock treatment in Eulerian fluid simulations, except for shock tracking \cite{rawat2010high}.
When letting $\sqrt{\alpha} \propto \Delta x \rightarrow 0$ jointly, we again observe second-order convergence before shock formation and first-order afterward.
In light of the results above, the observed convergence rate is consistent with the theoretical convergence rate. 

\subsection{Example cases}

We consider representative 1D--3D test problems that highlight IGR's behavior in the presence of shocks and fine-scale wave content.
For 1D smooth-wave propagation, we consider a sinusoidal wave benchmark.
For shock-dominated flows, we consider the Shu--Osher interaction problem and shock-tube benchmarks.
In 2D, we consider a Riemann interaction with added entropy-wave content and the Double Mach reflection problem.
In 3D, we consider compressible Taylor--Green turbulence and a multi-jet high-Mach configuration.
For baseline comparisons, we use fifth-order WENO reconstruction in two variants.
Characteristic WENO projects the conserved variables into the local characteristic fields of the flux Jacobian before applying the WENO stencil selection.
We combine it with a Roe Riemann solver \cite{toro2009riemann}. 
Component WENO applies WENO independently to each conserved variable. 
We combine it with a Rusanov (local Lax-Friedrichs) Riemann solver.
The characteristic decomposition typically reduces spurious oscillations near discontinuities, and the Roe approximate Riemann solver is less diffusive than the Rusanov one, but they entail higher computational costs.
\citet{wilfong2025simulating} report a fourfold speedup using IGR and replacing WENO with linear reconstruction of the same order and HLLC with Rusanov. 

In many applications, it is most convenient to smooth the initial conditions (see \cref{sec:smoothing}) via diffusion. 
For the sake of having a well-defined and grid-independent initial condition, we instead replace jumps by scaled hyperbolic tangent-like functions, which we refer to as $\mathrm{tanh}$ smoothing.

\subsubsection{1D smooth wave propagation}

To assess numerical dissipation in smooth flow, we consider the propagation of a small-amplitude sinusoidal velocity perturbation with periodic boundary conditions on $[0,1]$.
The initial condition is $\rho = 1$, $u = \beta \sin(2\pi k x)$ with amplitude $\beta = 0.01$ and wavenumber $k = 20$, and $p = 1.6$ with $\gamma = 1.4$.
The final time is $T = 0.4$.

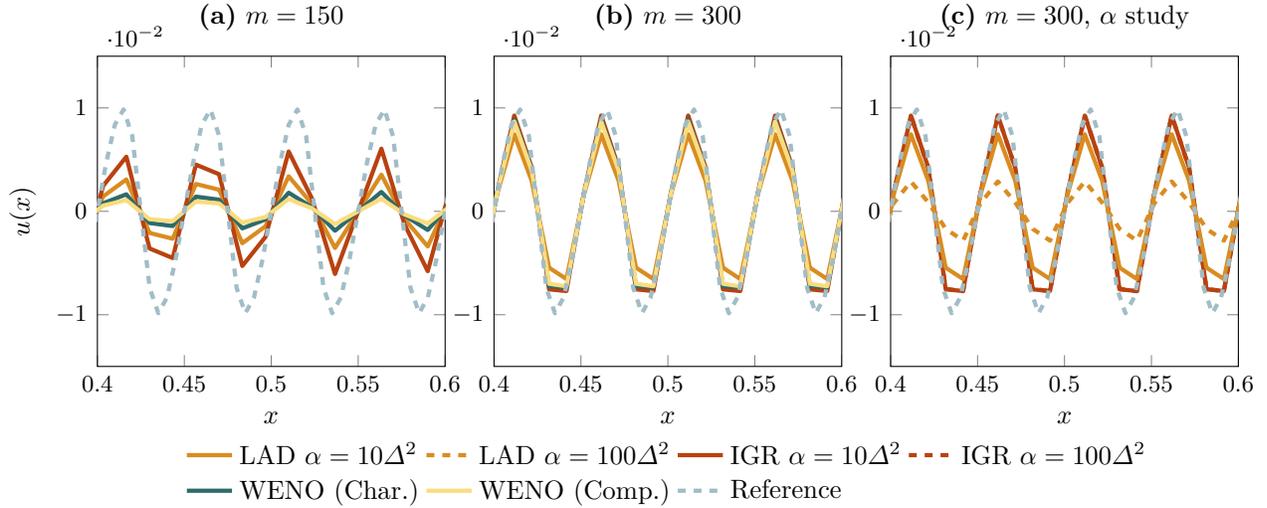
\begin{figure}[tbhp]
    \centering
    \tikzsetnextfilename{sine_wave}
    \begin{tikzpicture}\end{tikzpicture}
    \caption{
        Sine wave propagation comparing velocity profiles at final time $T = 0.4$.
        \textbf{(a)} and~\textbf{(b)} compare all methods at $m = 150$ and $m = 300$, respectively.
        \textbf{(c)}~shows IGR and LAD at $m = 300$ with $\alpha = 10\Delta x^2$ and $100\Delta x^2$.
        The reference solution (dashed blue) is computed with characteristic WENO and $m = 10^4$.
    }
    \label{fig:sine_wave}
\end{figure}

We compare IGR ($\alpha = 2\Delta^2$), LAD, characteristic WENO, and component WENO at resolutions $m = 150$ and $m = 300$, using a characteristic WENO solution at $m = 10^4$ as the reference.
Results are shown in \cref{fig:sine_wave}: panels~(a) and~(b) compare all methods at $m = 150$ and $m = 300$, respectively; panel~(c) shows the effect of increasing $\alpha$ on both IGR and LAD at $m = 300$, with $\alpha = 10\Delta^2$ and $100\Delta^2$.
At the default $\alpha = 2\Delta^2$ and $m = 150$, IGR closely tracks the reference solution and is significantly less dissipative than either characteristic WENO or LAD.
At $m = 300$, IGR achieves accuracy comparable to characteristic WENO, while LAD remains visibly more dissipative.
In panel~(c), increasing $\alpha$ from $10\Delta^2$ to $100\Delta^2$ has a negligible effect on IGR, whereas LAD's amplitude at $\alpha = 100\Delta^2$ drops to roughly one-third of its default value.
This asymmetry shows that IGR is notably more robust to the choice of $\alpha$ than LAD.

\subsubsection{1D Shu--Osher interaction}

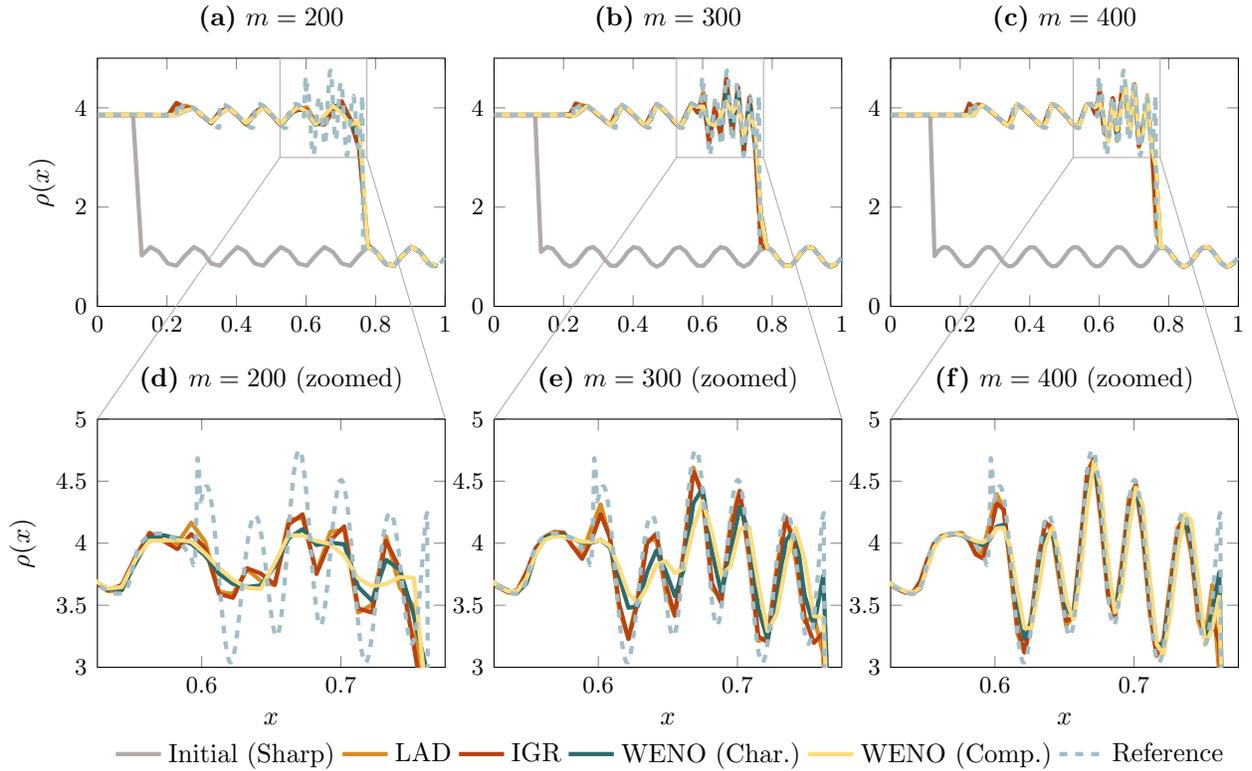
\begin{figure}[tbhp]
    \centering
    \tikzsetnextfilename{shu_osher}
    \begin{tikzpicture}\end{tikzpicture}
    \caption{
        Shu--Osher problem comparing density profiles at resolutions $m = 200$, $300$, and $400$. The top row~\textbf{(a--c)} shows the full domain; the bottom row~\textbf{(d--f)} shows the post-shock oscillatory region.
        All methods are compared against a reference solution computed with characteristic WENO at $m = 5000$.
    }
    \label{fig:shu_osher}
\end{figure}

The Shu--Osher problem~\cite{shu1988efficient} models the interaction of a Mach~3 shock with a sinusoidal entropy perturbation (at constant pressure).
We solve a rescaled version on the domain $[0,1]$ with $\gamma = 1.4$ and final time $t = 0.18$.
The post-shock state is $(\rho, u, p) = (27/7, 4\sqrt{35}/9, 31/3)$ and the pre-shock region has density $\rho = 1 + 0.2\sin(16\pi x)$ with $u = 0$ and $p = 1$.
We compare IGR ($\alpha = 2\Delta^2$) with LAD, characteristic WENO, and component-wise WENO at resolutions $m = 200$, $300$, and $400$.
The reference solution is computed with characteristic WENO at $m = 5000$.
The initial condition is $\mathrm{tanh}$ smoothed for IGR and LAD; WENO uses the sharp initial condition directly.
Results are shown in \cref{fig:shu_osher}: the top row shows the full density profile, and the bottom row magnifies the post-shock oscillatory region.
In the oscillatory region, IGR and LAD preserve the high-frequency density fluctuations more accurately than the WENO methods.

\subsubsection{1D Sod shock tube}

The Sod shock tube~\cite{sod1978survey} is a standard Riemann problem that produces a right-traveling shock, a contact discontinuity, and a left-traveling rarefaction.
We solve it on the domain $[0,1]$ with $\gamma = 1.4$ and final time $t = 0.2$.
The initial condition is $(\rho, u, p) = (1, 0, 1)$ for $x < 0.5$ and $(\rho, u, p) = (0.125, 0, 0.1)$ for $x \geq 0.5$.

We compare IGR ($\alpha = 2\Delta^2$) with LAD, characteristic WENO, and component WENO at resolutions $m = 200$, $400$, and $800$.
The initial condition is $\mathrm{tanh}$ smoothed for IGR and LAD with smoothing parameter $\varepsilon = 4/m$, so that $\varepsilon$ halves as $m$ doubles, mirroring the $\sqrt{\alpha} \propto \Delta x$ scaling from \cref{fig:igr_convergence}.
WENO uses the sharp initial condition directly.
All results are compared against the exact Riemann solver solution.

\begin{figure}[tbhp]
    \centering
    \tikzsetnextfilename{sod_shock}
    \begin{tikzpicture}\end{tikzpicture}
    \caption{
        Sod shock tube convergence study.
        Rows~1--2~\textbf{(a--f)}: density profiles, full and zoomed near the contact discontinuity.
        Rows~3--4~\textbf{(g--l)}: specific internal energy profiles, full and zoomed.
        Columns correspond to $(m, \varepsilon)$ = $(200, 0.02)$, $(400, 0.01)$, and $(800, 0.005)$.
        WENO uses the sharp initial condition; LAD and IGR use smoothed initial conditions.
        All results are compared against the exact Riemann solver solution.}
    \label{fig:sod}
\end{figure}
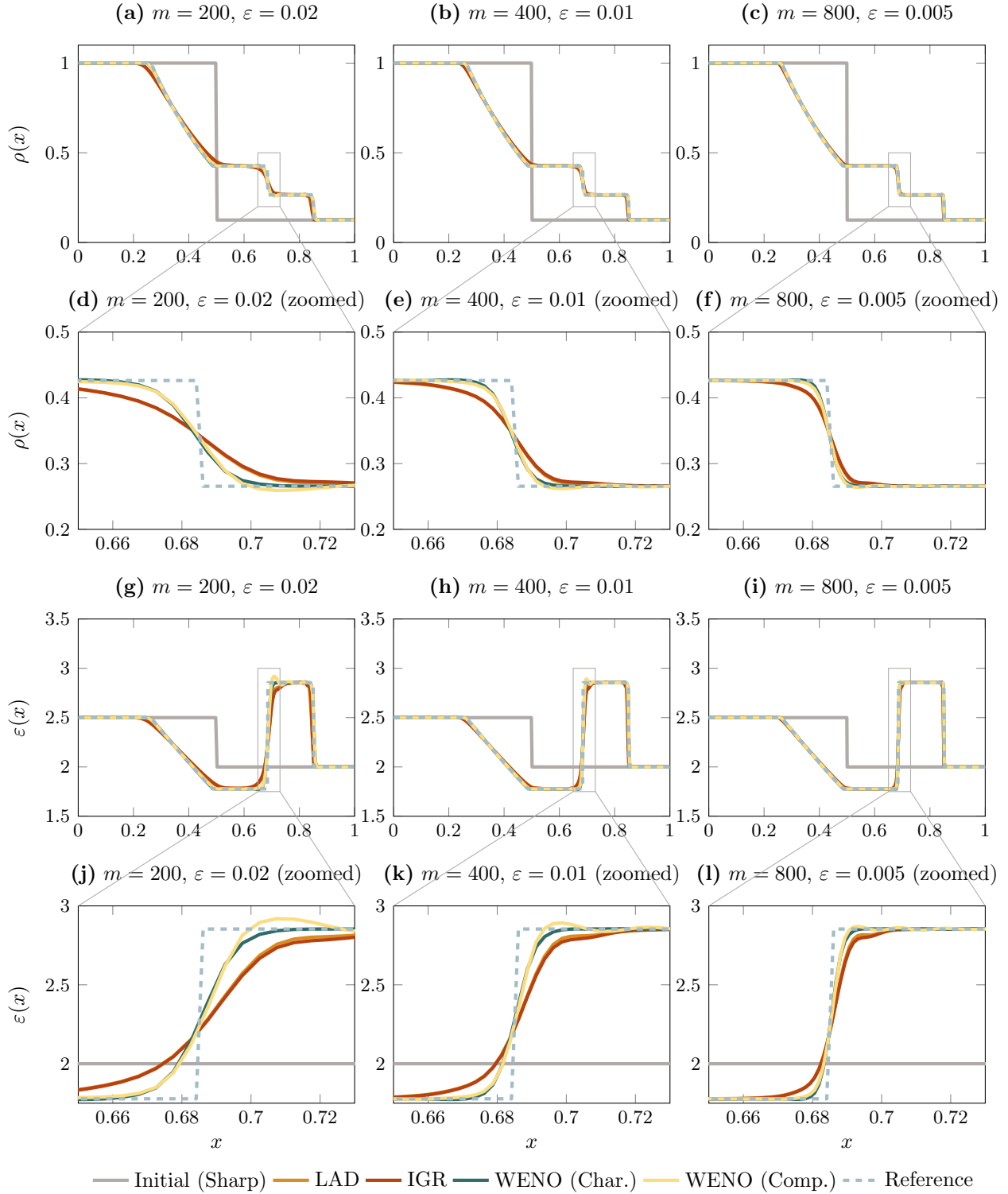

\Cref{fig:sod} is organized as a $3 \times 4$ grid: the first two rows show density profiles (full and zoomed near the contact discontinuity), and the last two rows show specific internal energy profiles (full and zoomed).
The zoomed panels focus on the contact discontinuity region ($x \in [0.65, 0.73]$), where differences between methods are most apparent.
All methods converge at approximately first order, consistent with the presence of discontinuities.
Near the contact, characteristic WENO achieves a sharper profile than IGR or LAD, owing to its use of the sharp initial condition.
IGR and LAD produce profiles of similar width, set by the smoothing parameter $\varepsilon \propto \Delta x$ of the initial condition rather than by numerical diffusion from the LF solver; the contact sharpens at the same rate as $\varepsilon$ as resolution increases.

\subsubsection{1D Leblanc shock tube}

The Leblanc shock tube~\cite{loubere2005subcell} features a large density ratio ($10^3$) and pressure ratio (${\sim}10^{8}$), making it a stringent test of robustness under large jumps of the state variables.

We solve this problem on the domain $[0,9]$ with $\gamma = 5/3$ and final time $t = 6$.
The left (high-energy) state $(\rho, u, p) = (1, 0, 1/15)$ is initialized in a single cell at the left boundary ($x_0 = 0$), and the ambient right state $(\rho, u, p) = (10^{-3}, 0, (2/3)\times 10^{-9})$ fills the remainder of the domain.

We compare IGR ($\alpha = 2\Delta^2$), LAD, and characteristic WENO with $m = 450$, $900$, and $1800$ grid points across the 1D domain.
All results are compared against the exact Riemann solver solution computed from the sharp initial condition.
IGR and LAD require smoothed initial conditions and are thus only shown for $\varepsilon > 0$; WENO uses the sharp initial condition and appears in all panels.
Here, $\varepsilon$ is a dimensionless parameter controlling the width of the $\mathrm{tanh}$ smoothing the initial discontinuity.

\begin{figure}[tbhp]
    \centering
    \tikzsetnextfilename{leblanc_shock}
    \begin{tikzpicture}\end{tikzpicture}
    \caption{
        Leblanc shock tube convergence study showing specific internal energy profiles.
        Each row corresponds to a different smoothing parameter ($\varepsilon = 0, 0.025, 0.05, 0.1$) and each column shows a different resolution ($m = 450, 900, 1800$).
        Runs not shown amount to simulations that break down due to negative densities or energies.
        All results are compared against the exact Riemann solver solution for the sharp initial condition.
        For $\varepsilon = 0$, only WENO is shown, as IGR and LAD require smoothed initial conditions. Component-wise WENO broke down in all of the above settings.}
    \label{fig:leblanc}
\end{figure}
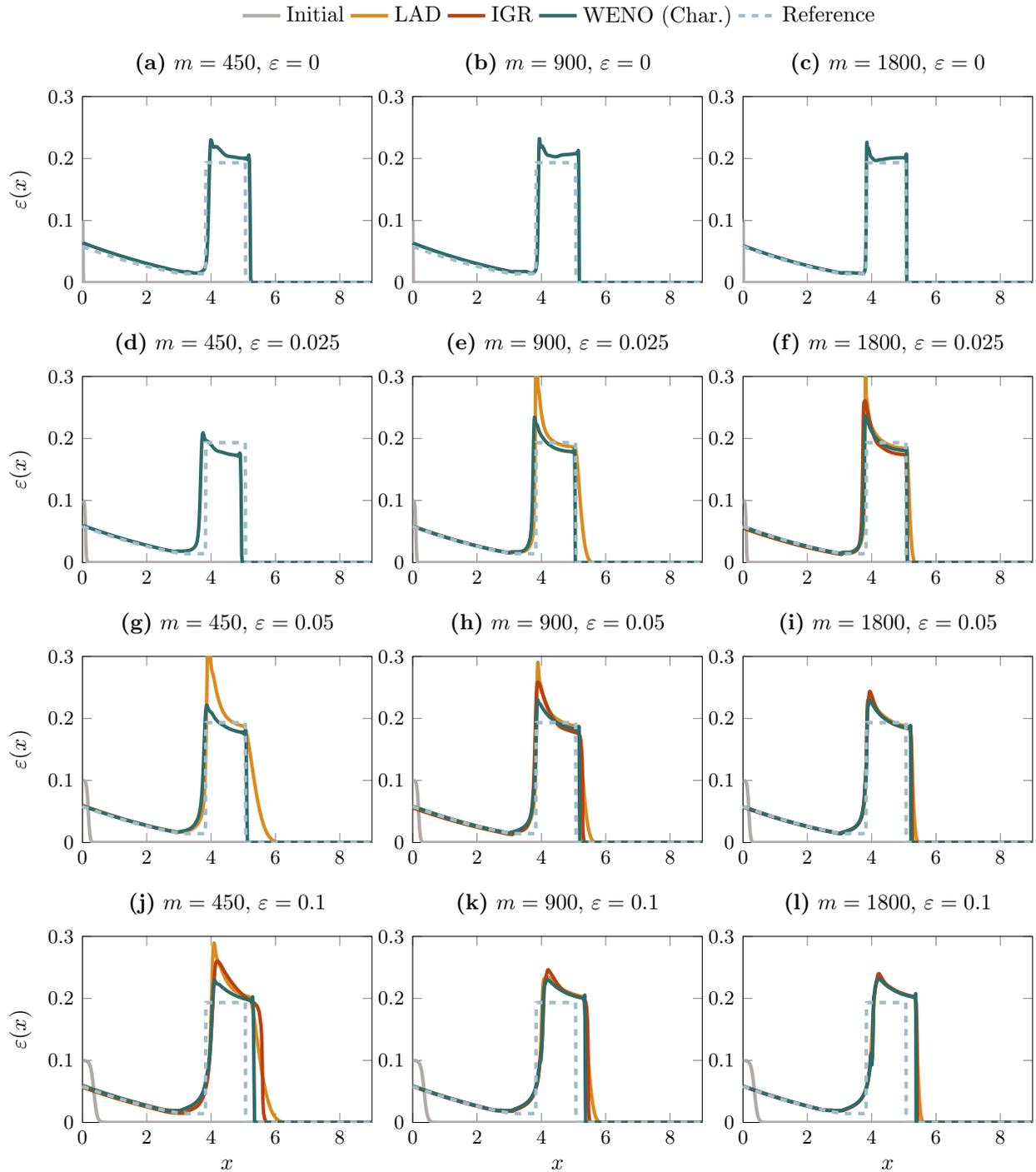

\Cref{fig:leblanc} presents a convergence study in specific internal energy, organized as a grid with rows corresponding to different smoothing parameters ($\varepsilon = 0, 0.025, 0.05, 0.1$, amounting to different degrees of regularization of the initial condition) and columns showing different resolutions ($m = 450, 900, 1800$).
As $\varepsilon$ decreases and resolution increases, the IGR and LAD solutions converge toward the exact solution, demonstrating that the smoothed-IC approach recovers the correct entropy solution in the limit.
However, IGR and LAD require progressively finer grid resolutions to avoid encountering negative densities or energies as the smoothing is decreased.
Component-wise WENO crashes at each fixed $\varepsilon$ and resolution, whereas Characteristic WENO achieves lower error than both IGR and LAD.
The relative performance of IGR and LAD is close across all configurations, with IGR showing marginally lower errors in most cases.

\subsubsection{2D Riemann interaction with entropy-wave content}

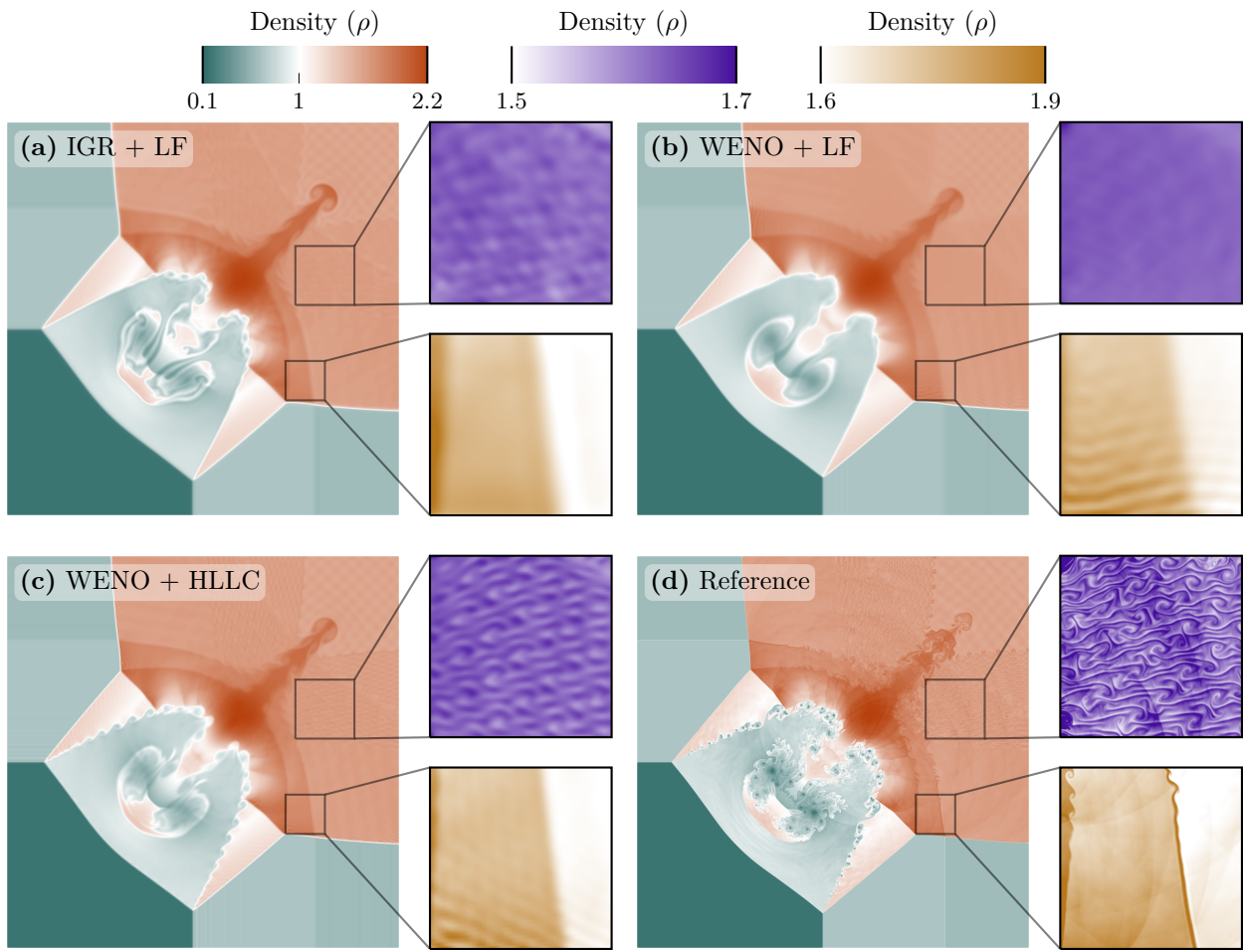
\begin{figure}[tbhp]
    \centering
    \tikzsetnextfilename{riemann_test_colorbar}
    \begin{tikzpicture}\end{tikzpicture}
    \\
    \begin{subfigure}{0.49\textwidth}
        \centering
        \tikzsetnextfilename{riemann_test_igr}
        \begin{tikzpicture}\end{tikzpicture} 
        \label{fig:riemann_igr}
    \end{subfigure}\hfill
    \begin{subfigure}{0.49\textwidth}
        \centering
        \tikzsetnextfilename{riemann_test_lf}
        \begin{tikzpicture}\end{tikzpicture}
        \label{fig:riemann_weno_lf}
    \end{subfigure}
    \begin{subfigure}{0.49\textwidth}
        \centering
        \tikzsetnextfilename{riemann_test_hllc}
        \begin{tikzpicture}\end{tikzpicture}
        \label{fig:riemann_weno_hllc}
    \end{subfigure}\hfill
    \begin{subfigure}{0.49\textwidth}
        \centering
        \tikzsetnextfilename{riemann_test_reference}
        \begin{tikzpicture}\end{tikzpicture}
        \label{fig:riemann_weno_ref}
    \end{subfigure}
    \caption{Riemann test problem with $N = 500^2$ grid points using \textbf{(a)} IGR/LF, \textbf{(b)} WENO5/LF, and \textbf{(c)} WENO5/HLLC with a \textbf{(d)} reference vanishing viscosity solution using WENO5/LF at $N = 5000^2$ grid points.} 
    \label{fig:riemann_test}
\end{figure}

We consider a 2D Riemann problem~\cite{liang2024new} on a computational domain with sizes $\left[0,1\right] \times \left[0,1\right]$ and a grid resolution $N = 500^2$.
The density, velocities, and pressure ($\rho, u, v, p$) in all regions of the computational domain are given in \cref{e:riemann_config}.

\begin{equation}
    \label{e:riemann_config}
    (\rho,u,v,p) = \begin{cases}
      (0.138,1.206,1.206,0.029) & x\leq0.75,\, y\leq0.75\\
      (0.532,0.000,1.206,0.300) & x\geq0.75,\, y\leq0.75\\
      (0.532,1.206,0.000,0.300) & x\leq0.75,\, y\geq0.75\\
      (1.500,0.000,0.000,1.500) & x\geq0.75,\, y\geq0.75
    \end{cases}
\end{equation}
An entropy wave is then introduced to this problem by adding fine-scale sinusoidal fluctuations in density (and consequently energy), while maintaining constant pressure. 
The resulting perturbed Riemann problem is run with IGR at $\alpha = 2\Delta^2$, where $\Delta$ is the grid spacing. 
For baseline comparisons, we use a 5th-order accurate WENO scheme with both the HLLC and Lax--Friedrichs (LF) Riemann solvers. 
We also run a well-resolved simulation at $N = 5000^2$ using WENO reconstruction and the LF Riemann solver, which we treat as the true solution in the vanishing viscosity limit. 
We apply $\mathrm{tanh}$ smoothing to the Riemann problem in \cref{e:riemann_config}, which is necessary for a stable IGR solution. 
\Cref{fig:riemann_test} shows density heat maps for all numerical methods considered.
Compared to the reference solution in \cref{fig:riemann_test}~(d), the flow structures match most closely with WENO5/HLLC (\cref{fig:riemann_test}~(c)), followed by IGR (\cref{fig:riemann_test}~(a)), and WENO5/LF (\cref{fig:riemann_test}~(b)).
With the Lax--Friedrichs Riemann solver, WENO reconstruction completely dissipates the density fluctuations, while IGR preserves them. 
Also, WENO reconstruction introduces spurious grid-aligned artifacts near the shock boundary with both HLLC and LF Riemann solvers, which are absent with IGR. 

\subsubsection{2D Double Mach reflection}

The double Mach reflection problem~\cite{woodward1984numerical} is a canonical test case for compressible flows.
A planar shock passing through a reflecting wedge produces a shock reflection, yielding a self-similar flow parametrized by the Mach number and the slope angle. 

The double-Mach case is set up in the usual way.
We use a rectangular domain of size $\left[0,4\right] \times \left[0,1\right]$, with a Mach $10$ oblique shock angled at $\ang{60}$ and a reflecting wall at the bottom of the domain. 
The reflective wall extends from $x = 1/6$ to $x = 4$, with the region to the left of $x = 1/6$ assigned post-shock flow. 
The top boundary follows the motion of the Mach $10$ shock, with the shock starting at $x = 1/6$ at the bottom boundary. 
The left boundary has initial post-shock values, and the right wall has zero-gradient boundary conditions. 

We conduct our simulation on an $800\times200$ grid, using IGR with $\alpha = 2\Delta^2$, where $\Delta$ is the grid size. 
Numerical stability requires $\mathrm{tanh}$ smoothing of the initial condition. 
Due to the reflective boundary condition, a low-order accurate reconstruction of the variables is necessary to maintain stability; for this purpose, we use third-order interpolation. 
Although WENO and IGR schemes are unstable with fifth-order accurate reconstruction, IGR is still stable with an increase in $\alpha$ to $10\Delta^2$.

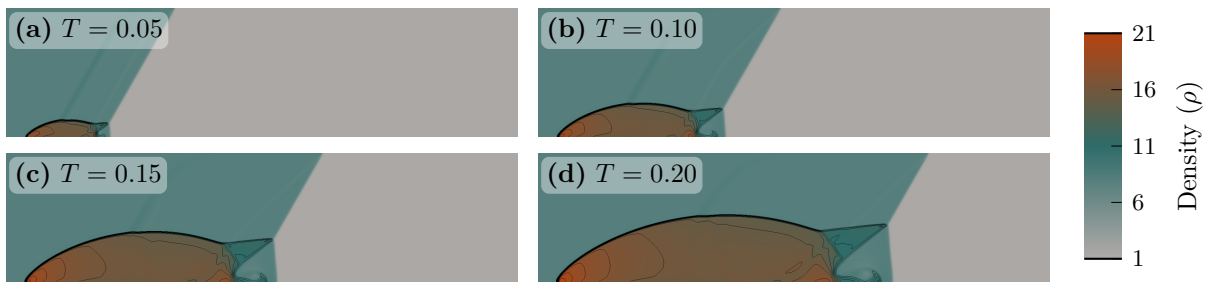
\begin{figure}[tbhp]
    \centering
    \begin{minipage}{0.84\textwidth}
        \begin{subfigure}{0.49\linewidth}
            \centering
            \tikzsetnextfilename{DM_1}
            \begin{tikzpicture}\end{tikzpicture}
            \label{fig:double_mach_1}
        \end{subfigure}\hfill
        \begin{subfigure}{0.49\linewidth}
            \centering
            \tikzsetnextfilename{DM_2}
            \begin{tikzpicture}\end{tikzpicture}
            \label{fig:double_mach_2}
        \end{subfigure}\vspace{0.5em}
        \begin{subfigure}{0.49\textwidth}
            \centering
            \tikzsetnextfilename{DM_3}
            \begin{tikzpicture}\end{tikzpicture}
            \label{fig:double_mach_3}
        \end{subfigure}\hfill
        \begin{subfigure}{0.49\textwidth}
            \centering
            \tikzsetnextfilename{DM_4}
            \begin{tikzpicture}\end{tikzpicture}
            \label{fig:double_mach_4}
        \end{subfigure}
    \end{minipage}\hfill
    \begin{minipage}{0.15\textwidth}
        \vspace{1.3em}
        \tikzsetnextfilename{double_mach_colorbar}
        \begin{tikzpicture}\end{tikzpicture} \\
    \end{minipage}
    \caption{
        Double Mach problem with $m_x \times m_y = 800 \times 200$ grid points at \textbf{(a)} $T = 0.05$, \textbf{(b)} $T = 0.10$, \textbf{(c)} $T = 0.15$, and \textbf{(d)} $T = 0.20$ using IGR and $\mathrm{tanh}$ smoothing.} 
    \label{fig:double_mach}
\end{figure}

The density contours at four simulation times with IGR are shown in \cref{fig:double_mach_1,fig:double_mach_2,fig:double_mach_3,fig:double_mach_4}.
The resulting flow structures match the expected results, except for an artifact originating at the shock's initial position, which becomes less prevalent as the simulation progresses and is largely absent at $T = 0.2$ in \cref{fig:double_mach_4}.
This artifact arises because the initial condition only approximately matches the profile of a fully developed IGR shock.  
As it forms an IGR shock at the onset of the simulation, a wave is reflected, causing the artifact.

\subsubsection{2D Isentropic vortex}

The isentropic vortex problem~\cite{shu2006essentially} is often used as a test problem because its exact solution enables evaluation of the accuracy and diffusion of numerical methods.

The vortex is initialized at the center of a periodic box of length $L = 10$. 
The initial conditions for density ($\rho$), velocity $(u,v)$, and pressure ($p$) are given by 
\begin{equation}
    \label{eq:isentropic}
    \begin{aligned}
    u &= u_{\infty} + y(S/2\pi) \exp\left(0.5\beta \left(1-r^2 \right) \right) \\
    v &= v_{\infty} - x(S/2\pi) \exp\left(0.5\beta \left(1-r^2 \right) \right) \\
    \rho &= \rho_\infty
    \left[
        1 - \frac{\rho_\infty}{p_\infty}
        \frac{S^2 (\gamma-1)}{8\gamma\pi^2}
        \exp\left( \beta \left(1-r^2 \right) \right)
    \right]^{\frac{1}{\gamma-1}} \\
    p &= p_\infty \left[
        1 - \frac{\rho_\infty}{p_\infty}
        \frac{S^2(\gamma-1)}{8\gamma\pi^2}
        \exp\left(\beta \left(1-r^2\right) \right)
    \right]^{\frac{\gamma}{\gamma-1}}
    \end{aligned}
\end{equation}
We use vortex strength $S = 5$, $\beta = 1$, and $\gamma = 1.4$ in \cref{eq:isentropic}, and $r^2 = x^2 + y^2$.
The free stream values for density ($\rho_\infty$) and pressure ($p_\infty$) are unity, and the $x$ and $y$ velocities are $u_\infty = 0.1$ and $v_\infty = 0$. 
The analytical solution for this problem advects the vortex at velocity $u_\infty = 0.1$ in the $x$ direction of the periodic domain. 
We evaluate the solution error after a fixed number of such periodic pass-through periods and compare the solution to the initial condition.

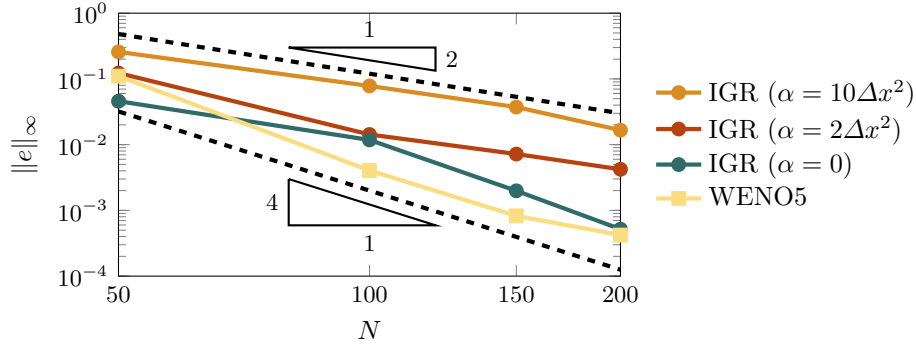
\begin{figure}[tbhp]
    \centering
    \tikzsetnextfilename{isentropic_vortex_error_norm}
    \begin{tikzpicture}\end{tikzpicture}
    \caption{Convergence rate for error $\|e\|_{\infty}$ in pressure with grid size for IGR with varying $\alpha$ and WENO5.}
    \label{fig:error_norm}
\end{figure}

We conduct an error convergence test using 5th-order accurate IGR and WENO schemes. 
IGR tests are conducted at varying values of $\alpha = (0,2,10) \Delta^2$, with $\Delta$ being the grid spacing.
The error $\|e\|_{\infty}$ in pressure $p$ after $4$ periods at varying grid sizes ($N = 50^2$, $100^2$, $150^2$, and $200^2$) is evaluated and plotted in \cref{fig:error_norm}.
Both the WENO5 and IGR with $\alpha$ = 0 exhibit 4th-order accurate error convergence in pressure, as expected. 
Due to the discretization of the right-hand side of the elliptic solve for $\Sigma$ (\cref{e:entropic_pressure}), IGR with $\alpha > 0$ can be at most 2nd-order accurate. 
The error grows with increasing $\alpha$ due to increased smoothing at a fixed resolution. 

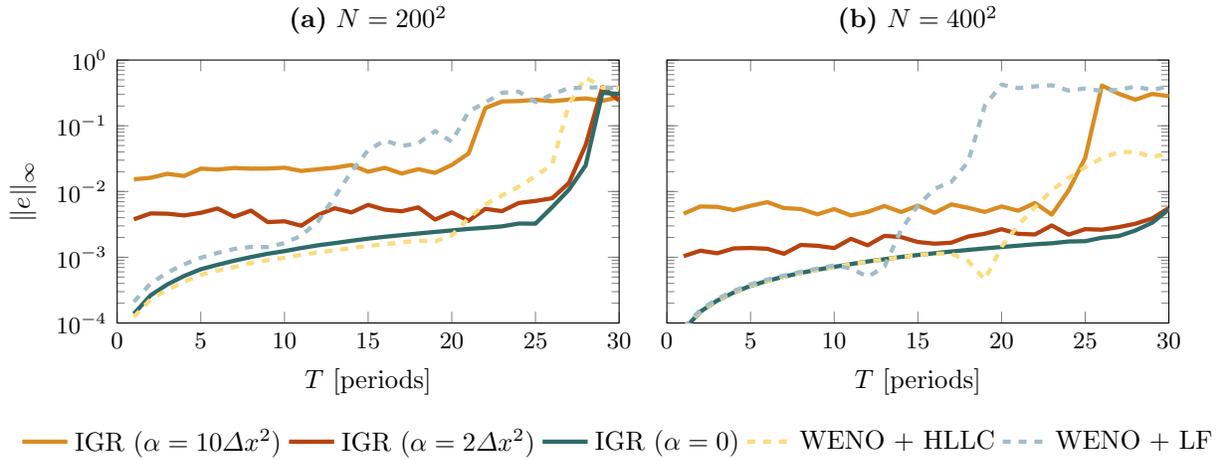
\begin{figure}[tbhp]
    \centering
    \tikzsetnextfilename{isentropic_vortex_error_growth}
    \begin{tikzpicture}\end{tikzpicture}
    \caption{Growth of error $\|e\|_{\infty}$ in pressure for an advecting isentropic vortex with grid sizes \textbf{(a)} $N=200^2$ and \textbf{(b)} $N=400^2$ versus the number of periods across the domain.}
    \label{fig:error_growth}
\end{figure}

To examine the effects of numerical dissipation, we investigate the error growth for the aforementioned schemes as the number of periods across the computational domain increases. 
We conduct WENO simulations with Lax--Friedrichs (LF) and HLLC Riemann solvers and IGR with LF. 
The test is conducted over $30$ periods at two different grid resolutions ((a)~$N = 200^2$ and (b)~$N = 400^2$) with the growth of error $\|e\|_{\infty}$ in pressure $p$ shown in \cref{fig:error_growth}. 

One sees that the error associated with projecting the initial condition stabilizes after sufficient $T$.
After the error becomes consistent, its growth rate exhibits two primary, well-established~\cite {spiegel2015survey} regimes: an initial linear growth phase and a later exponential one. 
The linear regime, which arises from numerical dissipation, is prominent for WENO (both Riemann solvers) and IGR with $\alpha = 0$.
In \cref{fig:error_growth}~(a), this lasts up to $T = 5$ and $20$ periods for WENO with LF and HLLC Riemann solvers, respectively, and up to $T = 25$ periods for IGR with $\alpha = 0$.
This regime is absent for IGR with $\alpha > 0$, with the error levels remaining constant due to lower numerical dissipation. 

All simulations eventually exhibit an exponential regime due to a destabilized vortex, leading to exponential error growth. 
This instability is independent of the numerical method and arises from the problem definition and its accompanying boundary conditions~\cite{spiegel2015survey}.
The onset of the exponential regime, however, occurs earlier for the WENO schemes: $T = 5$ and $20$ periods with LF and HLLC in \cref{fig:error_growth}~(a) 
We observe this onset for IGR at $T = 21$ and $25$ periods with $\alpha$ = $10\Delta^2$ and $2\Delta^2$ in \cref{fig:error_growth}~(a), with a smaller value of $\alpha$ further postponing it. 

\subsubsection{3D Taylor--Green vortex}

The 3D Taylor--Green vortex (TGV)~\cite{brachet1983small} tests IGR's ability to represent compressible turbulence. 
The TGV is set up in a 3D cube with side lengths $L = 2\pi$, with air at uniform density ($\rho = 1$) and the initial conditions for velocities $(u,v,w)$ and pressure $p$ given by \cref{eq:tgv}, where $p_0$ is the reference (ambient) pressure.
\begin{equation}
    \label{eq:tgv}
    \begin{aligned} 
        u &= \phantom{-}V_0\sin{(x)}\cos{(y)}\sin{(z)} \\
        v &= -V_0\cos{(x)}\sin{(y)}\sin{(z)} \\
        w &= 0 \\
        p &= p_0 + \frac{V_0^2}{16}(\cos{(2x)} + \cos{(2y)})(\cos{(2z)} + 2)
    \end{aligned}
\end{equation}
The Mach number $V_0/c_0$ is $1.25$, where $c_0 = \sqrt{\gamma p_0 / \rho}$ is the reference speed of sound, so shocklets appear in the flow~\cite{lusher2021assessment}. 
The Reynolds number is $\Rey = 10^4$ to have a sufficiently long inertial regime in the energy cascade, where the turbulent kinetic energy $E(k) \propto k^{-5/3}$. 
The inertial regime for this case typically ends at wave number $k = 2\pi/60\eta$~\cite{pope2001turbulent}, where $\eta \propto \Rey^{-3/4}$ is the Kolmogorov scaling. 
For $\Rey = 10^4$, this corresponds to an inertial regime that lasts up to $k = 100$.
We thus conduct simulations at $N = m_x \, m_y \, m_z = 128^3$, $256^3$, and $512^3$, which are sufficiently high to represent the inertial regime.

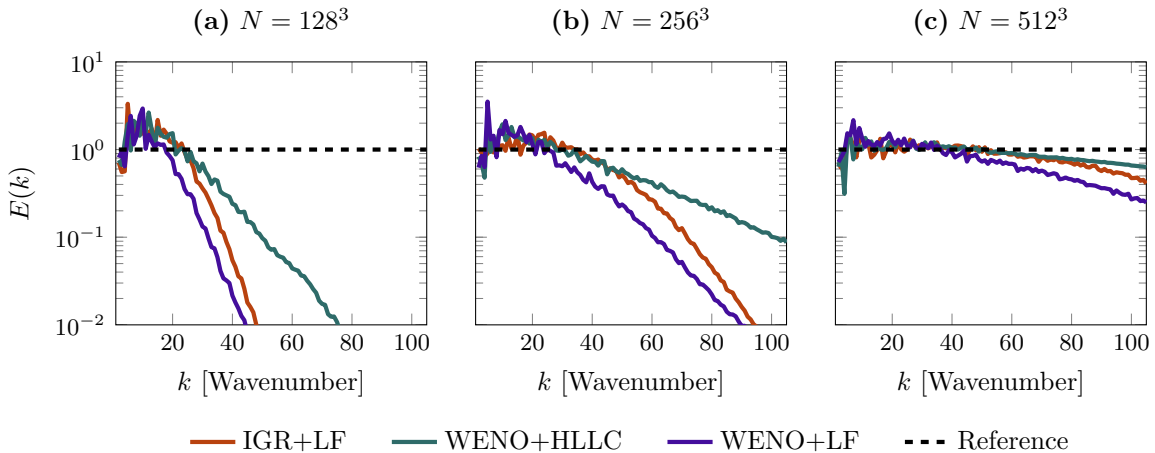
\begin{figure}[tbhp]
    \centering
    \tikzsetnextfilename{tgv_norm}
    \begin{tikzpicture}\end{tikzpicture}
    \caption{
        Turbulent kinetic energy cascade for IGR/LF, WENO5/HLLC, and WENO5/LF for $N = 128^3$, $256^3$, and $512^3$ grid points, normalized by a WENO5/HLLC solution that used $2048^3$ grid points.
    } 
    \label{fig:energy_cascade}
\end{figure}

We use IGR with $\alpha = 2\Delta^2$ and the Lax--Friedrichs (LF) Riemann solver, and compare the results to a baseline 5th-order accurate WENO (WENO5) scheme with HLLC and Lax--Friedrichs Riemann solvers. 
The turbulent energy cascade is then calculated using fast Fourier transforms and plotted at $t_c = 20L/V_0$.
The results are time-averaged across $t = 0.2t_c$ to minimize spurious oscillations. 
The energy cascade is normalized with a ground truth solution using WENO5 and HLLC at $N = 2048^3$ and plotted in \cref{fig:energy_cascade}. 
The results indicate notably greater dissipation with WENO than with IGR for the same Riemann solver (LF) and grid resolution, where IGR can sustain the inertial regime over a broader range of wavenumbers.
WENO reconstruction with HLLC, however, can improve upon IGR, with the gap narrowing at higher resolutions ($N = 512^3$).

\subsubsection{3D multi-jet configuration}

Many-engine rockets, such as the SpaceX Super Heavy, are powered by many smaller engines than the industry standard at the time of writing.
This choice comes with a key engineering challenge: high-speed exhaust plumes interact, leading to upstream-going waves and base heating~\cite{mehta2013numerical}. 

We simulate an array of $33$ rocket engines with an outlet velocity of Mach $10$ in atmospheric conditions, a configuration inspired by the SpaceX Super~Heavy. 
The simulation uses $600$ grid cells per engine outlet, resulting in $3.3 \times 10^{12}$ grid cells on a rectilinear grid.
This configuration was also used by the authors in the Gordon Bell work of~\citet{wilfong2025simulating}.

\begin{figure}[tbhp]
    \centering
    \tikzsetnextfilename{rockets}
    \begin{tikzpicture}\end{tikzpicture}
    \caption{
        Application of IGR to Mach 10 rocket simulations with strong shocks. 
        Panel (a) shows an artistic rendering of a 33-jet configuration. 
        Panels (b) and (c) show a comparison between IGR and WENO simulations of a simpler three-jet configuration.
        The IGR result in (b) avoids the grid-aligned artifacts present in the (c) WENO case.
        Images adapted from \citet{wilfong2025simulating} with author permission.
    }
    \label{fig:jets}
\end{figure}
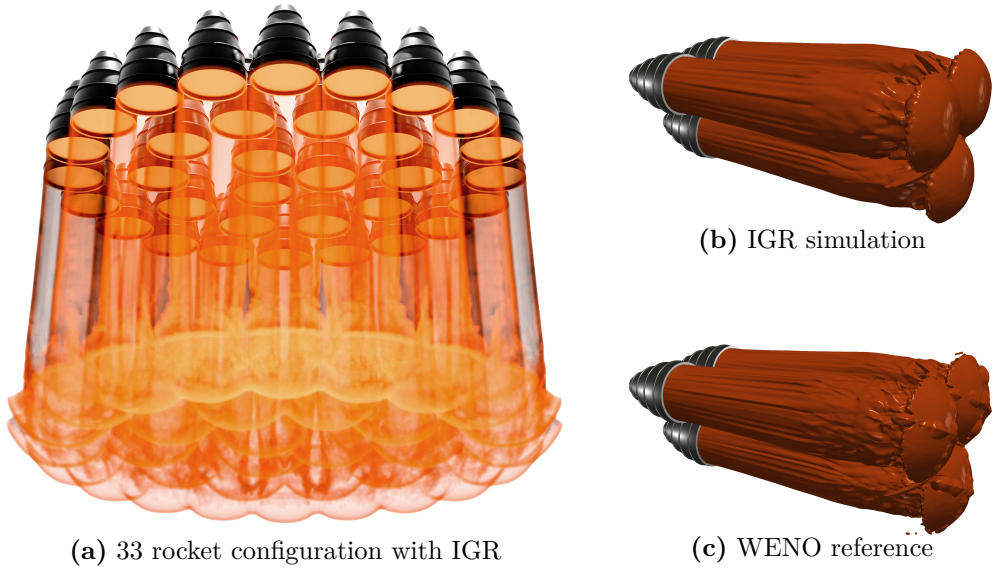

We use IGR with $\alpha = 2 \Delta^2$, where $\Delta$ is the grid spacing. 
We apply Gaussian smoothing to the initial condition to ensure numerical stability. 
The velocity contours for this simulation are shown in \cref{fig:jets}, and the fine-scale structures of interacting exhaust plumes are captured using IGR. 
A comparison of exhaust plumes for IGR and the baseline WENO simulation with HLLC Riemann solver for a simpler three-jet configuration is also presented in \cref{fig:jets}. 
Due to the baseline simulation's grid-dependent nature, grid-aligned artifacts are more prominent than with IGR. 

\section{Comparison to related work, discussion, and conclusions}\label{s:discussion}

\subsection{Comparison to related work}

Upwind finite-volume methods trace back to Godunov~\cite{godunov1959difference}, who solved local Riemann problems at cell interfaces to obtain numerical fluxes.
The piecewise parabolic method (PPM)~\cite{colella1984piecewise} and WENO~\cite{liu1994weighted,shu1998advanced} extended this to higher spatial order.
More complete Riemann solvers were developed to reduce interface dissipation, including the Roe~\cite{roe1981approximate}, HLLC~\cite{toro1994restoration,toro2009riemann}, and AUSM+~\cite{liou1996sequel} schemes.
High-order linear reconstruction near shocks introduces Gibbs oscillations; limiters such as MUSCL~\cite{vanleer1979towards,vanleer1974towards} and ENO~\cite{harten1987uniform} suppress these by adaptively reducing the approximation order near discontinuities.
The resulting mismatch between left- and right-interface states then determines the amount of dissipation added by the Riemann solver.
Flux-corrected transport (FCT)~\cite{boris1973flux,zalesak1979fully}, multi-dimensional optimal order detection (MOOD)~\cite{clain2011high}, and convex limiting~\cite{guermond2018second,kuzmin2020monolithic} take an a posteriori approach, applying corrections after the fact to maintain physical bounds.
Entropy-stable schemes~\cite{tadmor1987numerical} and central schemes~\cite{kurganov2000new} provide dissipation control without requiring a full Riemann solve.
The Runge--Kutta discontinuous Galerkin (DG) framework~\cite{cockburn1998runge} achieves high-order accuracy on unstructured meshes but still requires limiting or artificial viscosity near discontinuities.

Each of these approaches introduces nonlinear cell-to-cell coupling, thereby dissipating fine-scale features.
Approximate Riemann solvers such as Roe's scheme can additionally produce carbuncle instabilities on grid-aligned shocks~\cite{quirk1994contribution,chauvat2005shock}.
IGR produces smooth PDE solutions, so a linear Lax--Friedrichs flux suffices without a limiter or Riemann solver, avoiding these issues.

The 2D Riemann results show that IGR better preserves small-scale density fluctuations than WENO/LF, and avoids the grid-aligned artifacts present in both WENO/LF and WENO/HLLC.
The isentropic vortex results show that IGR has a lower error growth rate than WENO with either solver, reflecting reduced numerical dissipation in smooth flow.
The Taylor--Green vortex results show that IGR sustains the turbulent inertial range longer than WENO/LF at the same resolution.

Viscous regularizations, including artificial viscosity~\cite{vonneumann1950method}, entropy viscosity~\cite{guermond2011entropy}, LAD~\cite{cook2005hyperviscosity,mani2009suitability}, and PDE-based AV~\cite{barter2010shock}, smooth shocks through added dissipation, which also damps turbulence and acoustic waves.
IGR is an inviscid regularization; the Shu--Osher results show better preservation of high-frequency wave content than LAD, and the Taylor--Green vortex results show lower dissipation than WENO/LF at the same resolution.
Shock tracking~\cite{glimm1998three} avoids smearing by explicitly evolving the shock surface, recovering sharp profiles without diffusion.
This approach, however, is more complex to implement for multi-dimensional and topology-changing configurations.
IGR produces smooth shock profiles of controllable width $\sqrt{\alpha}$ within a standard finite-volume solver.

\subsection{Limitations of this work}

IGR requires initial conditions that are smooth on the scale $\sqrt{\alpha}$.
Sharp initial data must be preprocessed with a smoothing step, as done in several test cases here, but this has no physical consequence since IGR solutions converge to the correct entropy solution as $\alpha \to 0$.
This is a practical consideration when setting up a simulation.

The comparisons in this work also show that WENO5/HLLC can match or outperform IGR in certain regimes.
In the 2D Riemann test, WENO5/HLLC reproduces the reference solution most closely overall, and in the Taylor--Green vortex, the performance gap between IGR and WENO5/HLLC narrows at higher resolutions.
IGR's advantages are most pronounced when paired with a more dissipative (and cheaper) Riemann solver such as Lax--Friedrichs, or when fine-scale features must be preserved alongside shocks.
In this work, we use combine IGR with the linear Lax-Friedrichs numerical Riemann solver and the linear fifth-order reconstruction obtained by WENO5 in smooth regions.
However, we emphasize that the use of IGR is mostly agnostic to the choice of numerical method.
We believe that the combination of IGR with more accurate high-order discretizations, such as minimally dispersive methods developed for turbulence simulation \cite{lele1992compact}, holds great promise that we aim to explore in future work.

The convergence order of a numerical method is not necessarily reflective of its performance in the presence of localized features.  
More important is the cost--resolution tradeoff: ``how much computation is needed to resolve flow features of a given scale.'' 
For this goal, increasing the order of a numerical method is helpful only to a point \cite{rider2006effective}. 
The present work focuses on WENO5 due to its widespread use in the CFD community. 
We point out that alternatives such as piecewise linear (PLM) and piecewise quadratic (PPM) reconstructions were reported to yield better cost--accuracy tradeoffs by \cite{greenough2004quantitative}. 
This is subject to the caveat that the ever-widening gap between memory access and compute is pushes the needle toward higher order methods that are more arithmetically intense.
Nevertheless, we emhasize that IGR is agnostic to the underlying numerical method and that this work serves as a template for deploying IGR to many other numerical methods. 
Just like in the case of WENO, IGR could allow deploying PLM and PPM without slope limiters. 
This decreases computational cost as well as the diffusive effect of the approximate Riemann solver.
We defer an investigation of this setting to future work.
The under-resolved cases presented here, such as the rocket plume, operate in an implicit LES regime in which the numerical scheme and IGR together act as the effective filter.
The application of IGR in explicit LES using, for instance, a Smagorinsky-type subgrid model, is an interesting direction of future work.

We use $\alpha = 2\Delta^2$ throughout, but we set it empirically.
Rigorous guidance on the optimal choice of $\alpha$ for a given problem does not yet exist.

Because the entropic pressure $\Sigma$ is defined by a linear, self-adjoint elliptic operator, differentiating through the IGR system is structurally well-posed: the adjoint requires only a second elliptic solve of the same type.
Adjoint-based sensitivity computation with IGR has not been demonstrated in this work and is identified as a promising direction for future research.
The theoretical analysis of IGR convergence to entropy solutions has only been established for the 1D pressureless case~\cite{cao2024information}; extending these guarantees to the full compressible Navier--Stokes system remains open.

All test cases in this work use an ideal gas equation of state.
The behavior of IGR in multi-species, real-gas, or reactive thermodynamic systems has not been investigated.

\subsection{Conclusion}

This work demonstrates the utility of information geometric regularization (IGR) within a finite-volume framework for the compressible Navier--Stokes equations.
IGR replaces singular shocks with smooth profiles of user-controlled width $\sqrt{\alpha}$, eliminating the need for limiters and nonlinear Riemann solvers.
The resulting solver reduces to a linear reconstruction paired with a Lax--Friedrichs solve, lowering computational overhead and enabling kernel fusion relative to standard limiter-based approaches.

The 1--3D test suite spans shock tubes, the Shu--Osher interaction, a 2D Riemann problem with entropy-wave content, the double Mach reflection, and the Taylor--Green vortex.
Across all cases, IGR preserves fine-scale flow features that limiter-based and viscous methods dissipate, and does so with a simpler numerical scheme.
These properties suggest that IGR is a viable replacement for conventional shock-capturing in finite-volume codes, particularly in compressible flows where shocks coexist with fine-scale wave content.

\section*{Acknowledgments}

This work was partially supported by the Predictive Science Academic Alliance Program (PSAAP Award DE-NA0004261 - ``The Center for Information Geometric Mechanics and Optimization (CIGMO)'') managed by the NNSA (National Nuclear Security Administration) Office of Advanced Simulation.

FS gratefully acknowledges support from the Air Force Office of Scientific Research under award number FA9550-23-1-0668 (Information Geometric Regularization for Simulation and Optimization of Supersonic Flow) and from the Alfred P. Sloan Foundation via a Sloan Research Fellowship in Mathematics.

SHB acknowledges support for this work from NVIDIA Corporation and Advanced Micro Devices, Inc. (AMD) through hardware donations.

This research used resources of the Oak Ridge Leadership Computing Facility at the Oak Ridge National Laboratory, which is supported by the Office of Science of the U.S. Department of Energy under Contract No. DE-AC05-00OR22725, under allocation CFD154.

This work used Delta at the National Center for Supercomputing Applications through allocation PHY210084 from the Advanced Cyberinfrastructure Coordination Ecosystem: Services \& Support (ACCESS) program, which is supported by National Science Foundation grants \#2138259, \#2138286, \#2138307, \#2137603, and \#2138296.
This research is part of the Delta research computing project, which is supported by the National Science Foundation (award OCI~2005572), and the State of Illinois. Delta is a joint effort of the University of Illinois at Urbana--Champaign and its National Center for Supercomputing Applications.

We thank the anonymous reviewers for their valuable comments and suggestions that helped us to improve this manuscript.

\section*{Declaration of generative AI and AI-assisted technologies in the manuscript preparation process.}

During the preparation of this work, the author(s) used Claude Code to set up numerical experiments and tabulate results for figure generation.
Grammarly was used to refine some of the prose.
After using these tools/services, the author(s) reviewed and edited the content as needed and take(s) full responsibility for the content of the published article.

\bibliographystyle{bibsty}
\bibliography{main.bib}

@article{bruno2022fc,
  title        = {{FC}-based shock-dynamics solver with neural-network localized artificial-viscosity assignment},
  author       = {Bruno, Oscar P and Hesthaven, Jan S and Leibovici, Daniel V},
  year         = 2022,
  journal      = {Journal of Computational Physics: X},
  publisher    = {Elsevier},
  volume       = 15,
  pages        = 100110,
}

@article{quirk1994contribution,
  title        = {A contribution to the great {R}iemann solver debate},
  author       = {Quirk, James J.},
  year         = 1994,
  journal      = {International Journal for Numerical Methods in Fluids},
  publisher    = {Wiley},
  volume       = 18,
  number       = 6,
  pages        = {555--574},
}

@article{vonneumann1950method,
  title        = {A method for the numerical calculation of hydrodynamic shocks},
  author       = {VonNeumann, John and Richtmyer, Robert D},
  year         = 1950,
  journal      = {Journal of Applied Physics},
  volume       = 21,
  number       = 3,
  pages        = {232--237},
}

@article{liang2024new,
  title        = {A new high-order shock-capturing {TENO} scheme combined with skew-symmetric-splitting method for compressible gas dynamics and turbulence simulation},
  author       = {Liang, Tian and Fu, Lin},
  year         = 2024,
  journal      = {Computer Physics Communications},
  publisher    = {Elsevier},
  volume       = 302,
  pages        = 109236,
}

@article{liou1996sequel,
  title        = {A sequel to {AUSM}, {Part II: AUSM+}-up for all speeds},
  author       = {Liou, Meng-Sing},
  year         = 2006,
  journal      = {Journal of Computational Physics},
  volume       = 214,
  number       = 1,
  pages        = {137--170},
}

@inproceedings{spiegel2015survey,
  title        = {A survey of the isentropic {E}uler vortex problem using high-order methods},
  author       = {Spiegel, Seth C and Huynh, HT and DeBonis, James R},
  year         = 2015,
  booktitle    = {22nd {AIAA} computational fluid dynamics conference},
  pages        = 2444,
}

@article{bodony2022adjoint,
  title        = {Adjoint-based sensitivity of shock-laden flows},
  author       = {Bodony, DJ and Fikl, A},
  year         = 2022,
  journal      = {Proceedings of the 2022 {CTR} summer program},
}

@incollection{shu1998advanced,
  title        = {Essentially non-oscillatory and weighted essentially non-oscillatory schemes for hyperbolic conservation laws},
  author       = {Shu, Chi-Wang},
  year         = 1998,
  booktitle    = {Advanced Numerical Approximation of Nonlinear Hyperbolic Equations},
  publisher    = {Springer},
  editor       = {Cockburn, Bernardo and Johnson, Claes and Shu, Chi-Wang and Tadmor, Eitan},
  pages        = {325--432},
  series       = {Lecture Notes in Mathematics},
  volume       = 1697,
}

@article{fiorina2007artificial,
  title        = {An artificial nonlinear diffusivity method for supersonic reacting flows with shocks},
  author       = {Fiorina, Benoit and Lele, Sanjiva K},
  year         = 2007,
  journal      = {Journal of Computational Physics},
  publisher    = {Elsevier},
  volume       = 222,
  number       = 1,
  pages        = {246--264},
}

@article{roe1981approximate,
  title        = {Approximate {R}iemann solvers, parameter vectors, and difference schemes},
  author       = {Roe, Philip L},
  year         = 1981,
  journal      = {Journal of Computational Physics},
  publisher    = {Elsevier},
  volume       = 43,
  number       = 2,
  pages        = {357--372},
}

@article{lusher2021assessment,
  title        = {Assessment of low-dissipative shock-capturing schemes for the compressible {T}aylor--{G}reen vortex},
  author       = {Lusher, David J and Sandham, Neil D},
  year         = 2021,
  journal      = {AIAA Journal},
  publisher    = {American Institute of Aeronautics and Astronautics},
  volume       = 59,
  number       = 2,
  pages        = {533--545},
}

@article{guermond2011entropy,
  title        = {Entropy viscosity method for nonlinear conservation laws},
  author       = {Guermond, Jean-Luc and Pasquetti, Richard and Popov, Bojan},
  year         = 2011,
  journal      = {Journal of Computational Physics},
  publisher    = {Elsevier},
  volume       = 230,
  number       = 11,
  pages        = {4248--4267},
}

@incollection{shu2006essentially,
  title        = {Essentially non-oscillatory and weighted essentially non-oscillatory schemes for hyperbolic conservation laws},
  author       = {Shu, Chi-Wang},
  year         = 2006,
  booktitle    = {Advanced Numerical Approximation of Nonlinear Hyperbolic Equations: Lectures given at the 2nd Session of the Centro Internazionale Matematico Estivo ({CIME}) held in Cetraro, Italy, June 23--28, 1997},
  publisher    = {Springer},
  pages        = {325--432},
}

@inproceedings{koreeda1995front,
  title        = {Front structures of strong shock waves in air},
  author       = {Koreeda, J and Yanagisawa, H and Maeno, K and Honma, H and Bystrov, SA and Ivanov, VI and Shugaev, FV},
  year         = 1995,
  booktitle    = {Shock Waves {@} Marseille II: {P}hysico-Chemical Processes and Nonequilibrium Flow},
  pages        = {263--268},
  organization = {Springer},
}

@article{khesin2021geometric,
  title        = {Geometric hydrodynamics and infinite-dimensional Newton's equations},
  author       = {Khesin, Boris and Misio{\l}ek, Gerard and Modin, Klas},
  year         = 2021,
  journal      = {Bulletin of the American Mathematical Society},
  volume       = 58,
  number       = 3,
  pages        = {377--442},
}

@article{cook2005hyperviscosity,
  title        = {Hyperviscosity for shock-turbulence interactions},
  author       = {Cook, Andrew W and Cabot, William H},
  year         = 2005,
  journal      = {Journal of Computational Physics},
  publisher    = {Elsevier},
  volume       = 203,
  number       = 2,
  pages        = {379--385},
}

@article{cao2023information,
  title        = {Information geometric regularization of the barotropic {E}uler equation},
  author       = {Cao, Ruijia and Sch{\"a}fer, Florian},
  year         = 2023,
  journal      = {arXiv preprint arXiv:2308.14127},
}

@article{cao2024information,
  title        = {Information geometric regularization of unidimensional pressureless {E}uler equations yields global strong solutions},
  author       = {Cao, Ruijia and Sch{\"a}fer, Florian},
  year         = 2024,
  journal      = {arXiv preprint arXiv:2411.15121},
}

@article{mehta2013numerical,
  title        = {Numerical base heating sensitivity study for a four-rocket engine core configuration},
  author       = {Mehta, Manish and Canabal, Francisco and Tashakkor, Scott B and Smith, Sheldon D},
  year         = 2013,
  journal      = {Journal of Spacecraft and Rockets},
  publisher    = {American Institute of Aeronautics and Astronautics},
  volume       = 50,
  number       = 3,
  pages        = {509--526},
}

@article{puppo2004numerical,
  title        = {Numerical entropy production for central schemes},
  author       = {Puppo, Gabriella},
  year         = 2004,
  journal      = {SIAM Journal on Scientific Computing},
  publisher    = {SIAM},
  volume       = 25,
  number       = 4,
  pages        = {1382--1415},
}

@article{dolejvsi2003some,
  title        = {On some aspects of the discontinuous {G}alerkin finite element method for conservation laws},
  author       = {Dolej{\v{s}}{\'\i}, V{\'\i}t and Feistauer, Miloslav and Schwab, Christoph},
  year         = 2003,
  journal      = {Mathematics and Computers in Simulation},
  publisher    = {Elsevier},
  volume       = 61,
  number       = {3-6},
  pages        = {333--346},
}

@article{gottlieb1997gibbs,
  title        = {On the {G}ibbs phenomenon and its resolution},
  author       = {Gottlieb, David and Shu, Chi-Wang},
  year         = 1997,
  journal      = {SIAM Review},
  publisher    = {SIAM},
  volume       = 39,
  number       = 4,
  pages        = {644--668},
}

@article{harten1983upstream,
  title        = {On upstream differencing and {G}odunov-type schemes for hyperbolic conservation laws},
  author       = {Harten, Ami and Lax, Peter D. and {van Leer}, Bram},
  year         = 1983,
  journal      = {SIAM Review},
  volume       = 25,
  number       = 1,
  pages        = {35--61},
}

@article{toro1994restoration,
  title        = {Restoration of the contact surface in the {HLL}-{R}iemann solver},
  author       = {Toro, Eleuterio F. and Spruce, Matthew and Speares, William},
  year         = 1994,
  journal      = {Shock Waves},
  publisher    = {Springer},
  volume       = 4,
  number       = 1,
  pages        = {25--34},
}

@book{toro2009riemann,
  title        = {{R}iemann Solvers and Numerical Methods for Fluid Dynamics: {A} Practical Introduction},
  author       = {Toro, Eleuterio F.},
  year         = 2009,
  publisher    = {Springer},
  edition      = 3,
}

@article{barter2010shock,
  title        = {Shock capturing with {PDE}-based artificial viscosity for {DGFEM}: Part {I}. Formulation},
  author       = {Barter, Garrett E and Darmofal, David L},
  year         = 2010,
  journal      = {Journal of Computational Physics},
  publisher    = {Elsevier},
  volume       = 229,
  number       = 5,
  pages        = {1810--1827},
}

@inproceedings{wilfong2025simulating,
  title        = {Simulating many-engine spacecraft: {E}xceeding 1 quadrillion degrees of freedom via information geometric regularization},
  author       = {Wilfong, Benjamin and Radhakrishnan, Anand and {Le Berre}, Henry and Vickers, Daniel J. and Prathi, Tanush and Tselepidis, Nikolaos and Dorschner, Benedikt and Budiardja, Reuben and Cornille, Brian and Abbott, Stephen and Sch{\"a}fer, Florian and Bryngelson, Spencer H},
  year         = 2025,
  booktitle    = {Proceedings of {SC} '25: The International Conference for High Performance Computing, Networking, Storage and Analysis},
  pages       = {14--24},  
}

@article{brachet1983small,
  title        = {Small-scale structure of the {T}aylor--{G}reen vortex},
  author       = {Brachet, Marc E and Meiron, Daniel I and Orszag, Steven A and Nickel, Bernhard G and Morf, Rudolf H and Frisch, Uriel},
  year         = 1983,
  journal      = {Journal of Fluid Mechanics},
  publisher    = {Cambridge University Press},
  volume       = 130,
  pages        = {411--452},
}

@book{canuto2006spectral,
  title        = {Spectral Methods: {F}undamentals in Single Domains},
  author       = {Canuto, Claudio and Hussaini, M.Yousuff and Quarteroni, Alfio and Zang, Thomas A.},
  year         = 2006,
  publisher    = {Springer},
}

@article{gottlieb2001strong,
  title        = {Strong stability-preserving high-order time discretization methods},
  author       = {Gottlieb, Sigal and Shu, Chi-Wang and Tadmor, Eitan},
  year         = 2001,
  journal      = {SIAM Review},
  volume       = 43,
  number       = 1,
  pages        = {89--112},
}

@article{mani2009suitability,
  title        = {Suitability of artificial bulk viscosity for large-eddy simulation of turbulent flows with shocks},
  author       = {Mani, Ali and Larsson, Johan and Moin, Parviz},
  year         = 2009,
  journal      = {Journal of Computational Physics},
  volume       = 228,
  number       = 19,
  pages        = {7368--7374},
}

@article{arnold1966geometrie,
  title        = {Sur la g{\'e}om{\'e}trie diff{\'e}rentielle des groupes de {L}ie de dimension infinie et ses applications {\`a} l'hydrodynamique des fluides parfaits},
  author       = {Arnold, Vladimir},
  year         = 1966,
  journal      = {Annales de l'institut Fourier},
  volume       = 16,
  number       = 1,
  pages        = {319--361},
}

@article{woodward1984numerical,
  title        = {The numerical simulation of two-dimensional fluid flow with strong shocks},
  author       = {Woodward, Paul and Colella, Phillip},
  year         = 1984,
  journal      = {Journal of Computational Physics},
  volume       = 54,
  number       = 1,
  pages        = {115--173},
}

@article{vanleer1974towards,
  title        = {Towards the ultimate conservative difference scheme. {I. T}he quest of monotonicity},
  author       = {{van Leer}, Bram},
  year         = 1973,
  journal      = {Lecture Notes in Physics},
  volume       = 18,
  pages        = {163--168},
}

@article{vanleer1979towards,
  title        = {Towards the ultimate conservative difference scheme. {V.} {A} second-order sequel to {G}odunov's method},
  author       = {{van Leer}, Bram},
  year         = 1979,
  journal      = {Journal of Computational Physics},
  volume       = 32,
  number       = 1,
  pages        = {101--136},
}

@book{pope2001turbulent,
  title        = {Turbulent Flows},
  author       = {Pope, Stephen B.},
  year         = 2000,
  publisher    = {Cambridge University Press},
  address      = {Cambridge, UK},
}

@article{harten1987uniform,
  title        = {Uniformly high order accurate essentially non-oscillatory schemes, {III}},
  author       = {Harten, Ami and Engquist, Bj{\"o}rn and Osher, Stanley and Chakravarthy, S. R.},
  year         = 1987,
  journal      = {Journal of Computational Physics},
  publisher    = {Elsevier},
  volume       = 71,
  number       = 2,
  pages        = {231--303},
}

@article{lozano2019watch,
  title        = {Watch your adjoints! {L}ack of mesh convergence in inviscid adjoint solutions},
  author       = {Lozano, Carlos},
  year         = 2019,
  journal      = {AIAA Journal},
  publisher    = {American Institute of Aeronautics and Astronautics},
  volume       = 57,
  number       = 9,
  pages        = {3991--4006},
}

@article{liu1994weighted,
  title        = {Weighted essentially non-oscillatory schemes},
  author       = {Liu, Xi-Dong and Osher, Stanley and Chan, Tony},
  year         = 1994,
  journal      = {Journal of Computational Physics},
  publisher    = {Elsevier},
  volume       = 115,
  number       = 1,
  pages        = {200--212},
}

@article{barham2025hamiltonian,
  title={Hamiltonian Information Geometric Regularization of the Compressible Euler Equations},
  author={Barham, William and Tran, Brian K and Southworth, Ben S and Sch{\"a}fer, Florian},
  journal={arXiv preprint arXiv:2512.13948},
  year={2025}
}

@article{guelmame2022hamiltonian,
  title={Hamiltonian regularisation of the unidimensional barotropic Euler equations},
  author={Guelmame, Billel and Clamond, Didier and Junca, St{\'e}phane},
  journal={Nonlinear Analysis: Real World Applications},
  volume={64},
  pages={103455},
  year={2022},
  publisher={Elsevier}
}

@article{guermond2018second,
  title={Second-order invariant domain preserving approximation of the Euler equations using convex limiting},
  author={Guermond, Jean-Luc and Nazarov, Murtazo and Popov, Bojan and Tomas, Ignacio},
  journal={SIAM Journal on Scientific Computing},
  volume={40},
  number={5},
  pages={A3211--A3239},
  year={2018},
  publisher={SIAM}
}

@article{kuzmin2020monolithic,
  title={Monolithic convex limiting for continuous finite element discretizations of hyperbolic conservation laws},
  author={Kuzmin, Dmitri},
  journal={Computer Methods in Applied Mechanics and Engineering},
  volume={361},
  pages={112804},
  year={2020},
  publisher={Elsevier}
}

@article{chauvat2005shock,
  title={Shock wave numerical structure and the carbuncle phenomenon},
  author={Chauvat, Yann and Moschetta, J-M and Gressier, J{\'e}r{\'e}mie},
  journal={International Journal for Numerical Methods in Fluids},
  volume={47},
  number={8-9},
  pages={903--909},
  year={2005},
  publisher={Wiley Online Library}
}

@article{godunov1959difference,
  title        = {A difference method for numerical calculation of discontinuous solutions of the equations of hydrodynamics},
  author       = {Godunov, Sergei K.},
  year         = 1959,
  journal      = {Matematicheskii Sbornik},
  volume       = 47,
  number       = 3,
  pages        = {271--306},
}

@article{colella1984piecewise,
  title        = {The piecewise parabolic method ({PPM}) for gas-dynamical simulations},
  author       = {Colella, Phillip and Woodward, Paul R.},
  year         = 1984,
  journal      = {Journal of Computational Physics},
  publisher    = {Elsevier},
  volume       = 54,
  number       = 1,
  pages        = {174--201},
}

@article{cockburn1998runge,
  title        = {The {R}unge--{K}utta discontinuous {G}alerkin method for conservation laws {V}: Multidimensional systems},
  author       = {Cockburn, Bernardo and Shu, Chi-Wang},
  year         = 1998,
  journal      = {Journal of Computational Physics},
  publisher    = {Elsevier},
  volume       = 141,
  number       = 2,
  pages        = {199--224},
}

@article{tadmor1987numerical,
  title        = {The numerical viscosity of entropy stable schemes for systems of conservation laws. {I}},
  author       = {Tadmor, Eitan},
  year         = 1987,
  journal      = {Mathematics of Computation},
  volume       = 49,
  number       = 179,
  pages        = {91--103},
}

@article{kurganov2000new,
  title        = {New high-resolution central schemes for nonlinear conservation laws and convection--diffusion equations},
  author       = {Kurganov, Alexander and Tadmor, Eitan},
  year         = 2000,
  journal      = {Journal of Computational Physics},
  publisher    = {Elsevier},
  volume       = 160,
  number       = 1,
  pages        = {241--282},
}

@article{boris1973flux,
  title        = {Flux-corrected transport. {I. SHASTA}, a fluid transport algorithm that works},
  author       = {Boris, Jay P. and Book, David L.},
  year         = 1973,
  journal      = {Journal of Computational Physics},
  publisher    = {Elsevier},
  volume       = 11,
  number       = 1,
  pages        = {38--69},
}

@article{zalesak1979fully,
  title        = {Fully multidimensional flux-corrected transport algorithms for fluids},
  author       = {Zalesak, Steven T.},
  year         = 1979,
  journal      = {Journal of Computational Physics},
  publisher    = {Elsevier},
  volume       = 31,
  number       = 3,
  pages        = {335--362},
}

@article{glimm1998three,
  title        = {Three-dimensional front tracking},
  author       = {Glimm, James and Grove, John W. and Li, Xiaolin and Shyue, Keh-Ming and Zeng, Yanni and Zhang, Qiang},
  year         = 1998,
  journal      = {SIAM Journal on Scientific Computing},
  publisher    = {SIAM},
  volume       = 19,
  number       = 3,
  pages        = {703--727},
}

@article{clain2011high,
  title        = {A high-order finite volume method for systems of conservation laws---{M}ulti-dimensional {O}ptimal {O}rder {D}etection ({MOOD})},
  author       = {Clain, St{\'e}phane and Diot, Steve and Loub{\`e}re, Rapha{\"e}l},
  year         = 2011,
  journal      = {Journal of Computational Physics},
  publisher    = {Elsevier},
  volume       = 230,
  number       = 10,
  pages        = {4028--4050},
}

@article{shu1988efficient,
  title        = {Efficient implementation of essentially non-oscillatory shock-capturing schemes},
  author       = {Shu, Chi-Wang and Osher, Stanley},
  year         = 1988,
  journal      = {Journal of Computational Physics},
  publisher    = {Elsevier},
  volume       = 77,
  number       = 2,
  pages        = {439--471},
}

@article{sod1978survey,
  title        = {A survey of several finite difference methods for systems of nonlinear hyperbolic conservation laws},
  author       = {Sod, Gary A.},
  year         = 1978,
  journal      = {Journal of Computational Physics},
  publisher    = {Elsevier},
  volume       = 27,
  number       = 1,
  pages        = {1--31},
}

@article{harten1997uniformly,
  title={Uniformly high order accurate essentially non-oscillatory schemes, {III}},
  author={Harten, Ami and Engquist, Bjorn and Osher, Stanley and Chakravarthy, Sukumar R},
  journal={Journal of Computational Physics},
  volume={131},
  number={1},
  pages={3--47},
  year={1997},
  publisher={Elsevier}
}

@article{ray2018artificial,
  title={An artificial neural network as a troubled-cell indicator},
  author={Ray, Deep and Hesthaven, Jan S},
  journal={Journal of Computational Physics},
  volume={367},
  pages={166--191},
  year={2018},
  publisher={Elsevier}
}

@article{van1979towards,
  title={Towards the ultimate conservative difference scheme. {V}. A second-order sequel to {G}odunov's method},
  author={Van Leer, Bram},
  journal={Journal of Computational Physics},
  volume={32},
  number={1},
  pages={101--136},
  year={1979},
  publisher={Elsevier}
}

@article{glaubitz2019smooth,
  title        = {Smooth and compactly supported viscous sub-cell shock capturing for discontinuous {G}alerkin methods},
  author       = {Glaubitz, Jan and {\"O}ffner, Philipp and Ranocha, Hendrik and Sonar, Thomas},
  year         = 2019,
  journal      = {Journal of Scientific Computing},
  publisher    = {Springer},
  volume       = 79,
  number       = 1,
  pages        = {249--272},
}

@article{loubere2005subcell,
  title        = {A subcell remapping method on staggered polygonal grids for arbitrary {L}agrangian--{E}ulerian methods},
  author       = {Loub{\`e}re, Rapha{\"e}l and Shashkov, Mikhail J.},
  year         = 2005,
  journal      = {Journal of Computational Physics},
  publisher    = {Elsevier},
  volume       = 209,
  number       = 1,
  pages        = {105--138},
}

@article{rawat2010high,
  title={On high-order shock-fitting and front-tracking schemes for numerical simulation of shock--disturbance interactions},
  author={Rawat, Pradeep Singh and Zhong, Xiaolin},
  journal={Journal of Computational Physics},
  volume={229},
  number={19},
  pages={6744--6780},
  year={2010},
  publisher={Elsevier}
}

@article{lele1992compact,
  title={Compact finite difference schemes with spectral-like resolution},
  author={Lele, Sanjiva K},
  journal={Journal of Computational Physics},
  volume={103},
  number={1},
  pages={16--42},
  year={1992},
  publisher={Elsevier}
}

@inproceedings{nemec2006aerodynamic,
  title={Aerodynamic shape optimization using a Cartesian adjoint method and CAD geometry},
  author={Nemec, Marian and Aftosmis, Michael},
  booktitle={24th AIAA Applied Aerodynamics Conference},
  pages={3456},
  year={2006}
}

@inproceedings{rider2006effective,
  title={How effective are high-order approximations in shock-capturing methods? Is there a law of diminishing returns?},
  author={Rider, William J and Kamm, James R},
  booktitle={Computational Fluid Dynamics 2004: Proceedings of the Third International Conference on Computational Fluid Dynamics, ICCFD3, Toronto, 12--16 July 2004},
  pages={401--405},
  year={2006},
  organization={Springer}
}

@article{greenough2004quantitative,
  title={A quantitative comparison of numerical methods for the compressible Euler equations: fifth-order WENO and piecewise-linear Godunov},
  author={Greenough, JA and Rider, WJ},
  journal={Journal of Computational Physics},
  volume={196},
  number={1},
  pages={259--281},
  year={2004},
  publisher={Elsevier}
}
\end{document}